\documentclass[aip,jcp,graphicx,reprint,longbibliography,noeprint,citeautoscript]{revtex4-1}

\usepackage{comment}
\usepackage[pdftex]{graphics}
\usepackage{amssymb,amsmath}
\usepackage{siunitx}
\usepackage[colorlinks=true,linkcolor=black,citecolor=black,urlcolor=black]{hyperref}

\newcommand{\rev}[1]{\textcolor{black}{#1}} 

\begin{document}

\title{Economised path integrals}
\author{Zezhu Zeng}
\affiliation{Department of Chemistry, University of Oxford, Physical and Theoretical Chemistry Laboratory, South Parks Road, Oxford, OX1 3QZ, UK}
\author{David E. Manolopoulos}
\email{david.manolopoulos@chem.ox.ac.uk}
\affiliation{Department of Chemistry, University of Oxford, Physical and Theoretical Chemistry Laboratory, South Parks Road, Oxford, OX1 3QZ, UK}

\begin{abstract}
The Hessian of the ring polymer spring potential in the standard Trotter path integral is
a $P\times P$ symmetric circulant matrix with a centroid eigenvalue of zero. All such matrices commute and are diagonalised by the same bead to normal mode transformation matrix, and their eigenvalues contain $\lceil P/2\rceil-1$ degenerate pairs by symmetry. However, this still leaves some freedom to improve on the Trotter approximation: one can optimise the remaining $\lfloor P/2\rfloor$ independent non-zero normal mode frequencies to fit the exact quantum mechanical radii of gyration of harmonic ring polymers with frequencies in the range $0\le\omega\le\omega_{\rm max}$, where $\omega_{\rm max}$ is the maximum physical frequency in the problem of interest. The optimisation involves solving a simple least squares problem for the optimum (economised or “Eco”) internal mode frequencies. The remainder of the calculation then proceeds in the same way as a Trotter path integral calculation. An example application to hexagonal ice shows that the convergence of the Eco path integral is comparable to that of the 4th order Suzuki-Chin path integral, but with purely 2nd order Trotter effort. There is no need to calculate the projected Hessians that arise in the Suzuki-Chin method by finite differences, there is no need to develop any new estimators for observables, and once the Eco frequencies have been calculated the implementation of the Eco path integral involves changing just a few lines of a Trotter path integral code. To provide a more impressive example we have implemented the Eco method in GPUMD and used it to converge the (negative) thermal expansion coefficient and the constant pressure heat capacity of MOF-5 with a machine-learned neuroevolution potential. 

\end{abstract}

\maketitle

\section{Introduction}

The rapid development of machine-learned interatomic potentials for condensed phase systems since the pioneering papers by Behler and Parrinello\cite{Behler2007} and Csanyi and coworkers\cite{Bartok2010} has revolutionised the {\em ab initio} simulation of materials.  However, there are many situations in which an explicit treatment of nuclear quantum effects is needed in these simulations to obtain the correct result. Whereas older empirical potentials were fit to experimental data and therefore implicitly included some nuclear quantum effects, machine-learned {\em ab initio} potentials do not.  Classical molecular dynamics simulations using these potentials will therefore be unreliable for many systems, including almost all materials containing hydrogen atoms at room temperature and below.

The gold standard method for including nuclear quantum effects in materials simulations is the path integral molecular dynamics (PIMD) method of Parrinello and Rahman,\cite{Parrinello1984} which combines the isomorphism between the statistical mechanics of a quantum particle and that of a classical ring polymer\cite{Feynman1965,Chandler1981} with a molecular dynamics sampling scheme. This method is now well established, readily available in many computer programs, and widely used for materials simulations. However, the vast majority of modern PIMD simulations still use the primitive Trotter product discretisation of the imaginary time path, which  converges very slowly with respect to the number of ring polymer beads $P$. As a result, the calculation of quantum-sensitive properties such as the heat capacity can require hundreds of beads for full convergence,\cite{Markland2018} making the calculation orders of magnitude more expensive than a classical molecular dynamics simulation.

Various methods have been suggested over the years to reduce this expense. The oldest and most obvious approach is to replace the 2nd order Trotter discretisation with a more rapidly convergent path integral, such as one of the 4th order discretisations developed by Takahashi and Imada\cite{Takahashi1984} or Suzuki\cite{Suzuki1995} and Chin\cite{Chin1997}. These discretisations are straightforward to use in path integral Monte Carlo (PIMC) simulations,\cite{Takahashi1984} but they are more difficult to use in PIMD because the 4th order correction to the path integral action involves the derivative of the interaction potential. The forces it gives rise to therefore involve the Hessian of the potential, which is very expensive to compute in a large-scale condensed phase simulation. Kapil {\em et al.}\cite{Kapil2016} have recently overcome this difficulty by noting that one only requires the projection of the Hessian in a single direction, which can be evaluated more efficiently by finite differencing the force in that direction. Their projected Hessian Suzuki-Chin (SC) method is now well established and is probably the main competitor to the method we shall introduce in this paper. 

An alternative approach is to use a generalised Langevin equation to enforce the correct quantum fluctuations on a Trotter path integral in the harmonic limit, which can be done using a path integral generalised Langevin equation thermostat (PIGLET).\cite{Ceriotti2011,Ceriotti2012} While this can give significantly more accurate results than the basic Trotter path integral for small values of $P$, the convergence with increasing $P$ is no longer monotonic, and the method suffers from energy leakage from high to low frequency modes as the simulation proceeds. Here we shall show that the correct harmonic quantum fluctuations can be applied in a different way that avoids this energy leakage and leads to a genuine thermal equilibrium simulation rather than a non-equilibrium steady-state simulation of the type that PIGLET attempts to achieve.

Another interesting alternative is the perturbed path integral (PPI) approach of Poltavsky and Tkatchenko.\cite{Poltavsky2016} This regards the 4th order correction to the action in the Takahashi-Imada\cite{Takahashi1984} path integral as a perturbation, the effect of which on various observables can be accounted for by reweighting the results of a Trotter path integral simulation. This is an appealing idea because it only involves post-processing the results of an otherwise standard calculation. However, it has since been shown how the 4th order correction can be applied non-perturbatively with projected Hessians to give more rigorous 4th order estimators for observables,\cite{Kapil2016} so there is less need to make the perturbative approximation now than when it was first proposed.\cite{Poltavsky2016}

In this paper, we shall describe yet another approach to 
reducing $P$ in PIMD simulations. The inspiration for this approach comes from Hunt and Althorpe's recent paper on the Matsubara tail in the hierarchical equation of motion (HEOM) method for open system quantum dynamics.\cite{Hunt2026} By making new connections between the (Drude-Debye) bath correlation function in HEOM and the geometry of imaginary-time paths, Hunt and Althorpe showed how the ring polymer radius of gyration of a harmonic oscillator could be used to improve the treatment of the Matsubara tail. In particular, they showed how the Matsubara contribution could be represented more efficiently by fitting the squared radii of gyration $R^2(\omega)$ of harmonic oscillators with frequencies in the range $0\le\omega\le\omega_{\rm max}$ to the rational function 
\begin{equation*}
    R^2(\omega) \approx \sum_{n=1}^K \frac{k_n}{\omega^2+\eta_n^2}+k_0,
\end{equation*}
with suitably chosen poles $\pm i\eta_n$ and coefficients $k_n$.\cite{Hunt2026} We shall see in Eq.~(14) that a very similar fitting problem arises when the goal is to optimise the frequencies of the free ring polymer normal modes of a Trotter-type path integral to the same harmonic radius of gyration model, which explains the connection between their work and what we are about to propose.

Section~II presents the derivation of our new economised (or `Eco') path integral and explains why it can be used in the same way as a Trotter path integral -- with the same estimators for observables and with only very minor modifications to a standard Trotter path integral code.
Section~III presents an application of the Eco path integral to the kinetic and potential energies, radial distribution functions, and vibrational density of states of hexagonal ice at 100 K. Section~IV presents a more challenging application to the negative thermal expansion coefficient and the temperature dependence of the heat capacity of MOF-5. Section V summarises what we have accomplished and encourages the adoption of Eco for future PIMD simulations.

\section{Theory}

\subsection{The Trotter path integral}

It is straightforward to generalize any imaginary time path integral expression from one to more dimensions so it suffices to develop the present theory for a simple one-dimensional problem with a Hamiltonian of the form 
\begin{equation}
    \hat{H} = \frac{\hat{p}^2}{2m}+V(\hat{q}).
\end{equation}
The $P$-bead Trotter product approximation to the partition function of this problem, $Z = {\rm tr}\left[e^{-\beta\hat{H}}\right]$, is
\begin{equation}
Z_P = \frac{1}{(2\pi\hbar)^P} \int {\rm d}p\int {\rm d}q\, e^{-\beta_PH_P(p,q)},
\end{equation}
where $H_P(p,q)$ is the ring polymer Hamiltonian
\begin{equation}
H_P(p,q) = \sum_{j=0}^{P-1} \left[\frac{p_j^2}{2m}+\frac{1}{2}m\omega_P^2(q_j-q_{j+1})^2+V(q_j)\right],
\end{equation}
$\beta_P=\beta/P$, and $\omega_P=1/(\beta_P\hbar)$. All indices are assumed to be evaluated modulo $P$ both here and throughout what follows. 

The harmonic spring potential in Eq.~(3) can be written more succinctly as
\begin{equation}
\sum_{j=0}^{P-1} \frac{1}{2}m\omega_P^2(q_j-q_{j+1})^2 = \frac{1}{2} q^TKq \equiv {\sf U}(q).
\end{equation}
Here $K$ is a {\em real symmetric circulant matrix} with a centroid eigenvalue of zero:
\begin{equation}
K_{ij} = m\omega_P^2\,\kappa_{(i-j)\ \hbox{mod}\ P}
\end{equation}
where 
\begin{equation}
    \kappa_n = \kappa_{P-n}
\end{equation}
and
\begin{equation}
\sum_{n=0}^{P-1} \kappa_n = 0.
\end{equation}

Eq.~(5) is just the definition of a circulant matrix,\cite{Davis1994} in which we have separated the physical factor of $m\omega_P^2$ from the dimensionless circulant coefficients $\kappa_n$. This equation ensures that the ring polymer Hamiltonian $H_P(p,q)$ is invariant to a cyclic permutation of its beads ($j\to j+1$). Eq.~(6) ensures that $K$ is symmetric and that $H_P(p,q)$ is invariant to a reversal of the bead labels ($j\to P-j$). These properties of $H_P(p,q)$ stem from symmetries of the exact path integral expression for $Z$,\cite{Feynman1965}
\begin{equation}
    Z = \int {\cal D}q(\tau)\,\exp\left[-\frac{1}{\hbar}\int_{0}^{\beta\hbar} {\rm d}\tau \left\{\frac{m}{2}\dot{q}^2(\tau)+V(q(\tau))\right\}\right],
\end{equation}
which is unchanged by the substitutions $\tau'= (\tau+\delta\tau)\ \hbox{mod}\ \beta\hbar$ and $\tau'= \beta\hbar-\tau$. Eq.~(7) ensures that the centroid vector $c=(1,1,\ldots,1)^T/P$ is an eigenvector of $K$ with eigenvalue zero, or equivalently that the harmonic spring potential ${\sf U}(q)$ is independent of the ring polymer centroid coordinate $\bar{q}=c^Tq$.

The dimensionless circulant coefficients $\{\kappa_n\}_0^{P-1}$ of the Trotter path integral are $\kappa_0=2$, $\kappa_1=\kappa_{P-1}=-1$, and $\kappa_n=0$ for $n=2,\ldots,P-2$. These clearly satisfy Eqs.~(6) and~(7), but they are not the only circulant coefficients that do so. Since Eq.~(6) imposes $\lceil P/2\rceil -1$ constraints on the $P$ variables $\{\kappa_n\}_0^{P-1}$, and Eq.~(7) imposes one further constraint, there are still $P-\lceil P/2\rceil = \lfloor P/2\rfloor$ free parameters (e.g., $\kappa_1,\ldots,\kappa_{\lfloor P/2\rfloor}$) in a general $K$ matrix of the form in Eqs.~(5) to (7). We shall show below that  the existence of these free parameters can be exploited to improve the accuracy of the path integral approximation. However, since it is not so clear how to do this when working with the circulant coefficients $\{\kappa_n\}_1^{\lfloor P/2\rfloor}$ themselves, we shall first transform the problem to the normal mode representation (the eigenbasis of $K$).

All real symmetric $P\times P$ circulant $K$ matrices commute and are diagonalised by the same orthogonal bead-to-normal mode transformation matrix, the elements of which can be chosen to be\cite{Ceriotti2010}
\begin{equation}
C_{jk}= 
\begin{cases} 
\sqrt{\dfrac{1}{P}}, & k=0 \\[10pt]
\sqrt{\dfrac{2}{P}}\,\cos\!\left(\dfrac{2\pi jk}{P}\right), & 1\le k\le \left\lfloor\dfrac{P-1}{2}\right\rfloor \\[10pt]
\sqrt{\dfrac{1}{P}}\,(-1)^j, & P\ \hbox{even and}\ k=\dfrac{P}{2} \\[10pt]
\sqrt{\dfrac{2}{P}}\,\sin\!\left(\dfrac{2\pi jk}{P}\right), & \left\lceil\dfrac{P+1}{2}\right\rceil\le k\le P-1 
\end{cases}
\end{equation}
for $j,k=0,\ldots,P-1$. The corresponding eigenvalues $\lambda_k$ contain $\lceil P/2\rceil-1$ degenerate pairs by symmetry,
\begin{equation}
\lambda_k = \lambda_{P-k},
\end{equation}
and since the centroid normal mode has $\lambda_0=0$ this again leaves $\lfloor P/2\rfloor$ free parameters $\{\lambda_k\}_1^{\lfloor P/2\rfloor}$ with which to improve on the Trotter approximation.

If we identify $\lambda_k$ with $m\omega_k^2$, and assume without loss of generality that the $\omega_k$'s are non-negative (as they are in the 
Trotter discretisation where $\omega_k = 2\omega_P\,\sin(k\pi/P)$ for $k=0,\ldots,P-1$), the free parameters in this representation are $\omega_1,\ldots,\omega_{\lfloor P/2\rfloor}$. These are entirely analogous to the free parameters $\kappa_1,\ldots,\kappa_{\lfloor P/2\rfloor}$ in the bead representation, which can be recovered from the  $\omega_k$'s as
\begin{equation}
    \kappa_n = \frac{1}{P} \sum_{k=0}^{P-1} \left(\frac{\omega_k}{\omega_P}\right)^2\cos\,\left(\frac{2\pi kn}{P}\right).
\end{equation}
[Note in passing that the inverse of this transform,
\begin{equation}
\omega_k^2 = \omega_P^2\sum_{n=0}^{P-1} \kappa_n\,\cos\!\left(\frac{2\pi kn}{P}\right),
\end{equation}
gives
\begin{equation}
\omega_0^2 = \omega_P^2\sum_{n=0}^{P-1} \kappa_n,
\end{equation}
which explains why Eq.~(7) sets the centroid eigenvalue to zero.] When it comes to optimizing the path integral, the frequencies $\omega_1,\ldots,\omega_{\lfloor P/2\rfloor}$ are distinctly more convenient parameters than the circulant coefficients $\kappa_1,\ldots,\kappa_{\lfloor P/2\rfloor}$, because the $\omega_k$'s appear explicitly in the relevant objective function [see Eqs.~(14) and (15) below].

\subsection{The Eco path integral}

Since the finite $P$ Trotter frequencies $\omega_k=2\omega_P\sin(k\pi/P)$ do not give the correct (infinite $P$) ring polymer radius of gyration for a harmonic oscillator, we can use the simple harmonic oscillator model to improve on them. The natural way to do this is to follow Hunt and Althorpe\cite{Hunt2026} and look for new frequencies $\omega_k$ that give a better approximation to the correct squared radii of gyration
\begin{align}
R^2(\omega) &= \frac{1}{\beta m\omega^2}\left[\frac{\beta\hbar\omega}{2}\coth\!\left(\frac{\beta\hbar\omega}{2}\right)-1\right]\nonumber\\
&\approx \frac{1}{\beta m\omega^2}\sum_{k=1}^{P-1} \frac{\omega^2}{\omega^2+\omega_k^2}
\end{align}
and (therefore) the correct quantum energy shifts
\begin{align}
\Delta E(\omega) &= E^{\rm quantum}(\omega)-E^{\rm classical}(\omega)\nonumber\\ &= \frac{1}{\beta}\left[\frac{\beta\hbar\omega}{2}\coth\!\left(\frac{\beta\hbar\omega}{2}\right)-1\right]\nonumber\\
&\approx \frac{1}{\beta}\sum_{k=1}^{P-1} \frac{\omega^2}{\omega^2+\omega_k^2}
\end{align}
of harmonic oscillators $V(q)=\frac{1}{2}m\omega^2q^2$ with frequencies in the range we happen to be interested in (e.g., $0\le\omega\le\omega_{\rm max}=4000$ cm$^{-1}$ for ice and water).

To minimise the root-mean-square (rms) fractional error in $R^2(\omega)$, we begin by defining $x=\beta\hbar\omega$, $y_k=\beta\hbar\omega_k$, and
\begin{equation}
    f(x) = \frac{x^2}{(x/2)\coth(x/2)-1}.
\end{equation}
We then minimise 
\begin{equation}
    s(y) = \frac{1}{x_{\rm max}}\int_0^{x_{\rm max}}{\rm d}x\,\left|\,f(x)\sum_{k=1}^{P-1}\frac{1}{x^2+y_k^2}-1\,\right|^2
\end{equation}
with respect to $\{y_k\}_1^{\lfloor P/2 \rfloor}$, where $x_{\rm max}=\beta\hbar\omega_{\rm max}$ and $y_k=y_{P-k}$ for $k=\lfloor P/2\rfloor+1,\ldots,P-1$. The rms fractional error in $R^2(\omega)$ is then $\sqrt{s(y^*)}$, and the optimised frequencies are $\omega_k=y^*_k/(\beta\hbar)$, where $y^*$ is the value of $y$ at the minimum. The rms fractional error in $\Delta E(\omega)$ is also $\sqrt{s(y^*)}$ as a result of the similarity between Eqs.~(15) and~(16).

\rev{In practice we use a regularised Newton method  to minimize $s(y)$ subject to the constraint $0<y_1\le \cdots \le y_{\lfloor{P/2\rfloor}}$, starting from the Trotter initial guess $y_k=2P\sin(k\pi/P)$.  We have found this to lead to slightly fewer iterations than the Matsubara initial guess $y_k=2\pi k$. The objective function is non-convex, so the Newton procedure cannot in general be expected to find the global minimum from an arbitrary initial guess. In practice, however, we find that both the Trotter and Matsubara initial guesses converge to the same minimum for every $P$ in Fig.~1.} Our minimisation routine is provided in the SI for others to use and incorporate in their path integral codes. It takes less than a second of computer time to find the minimum in every case we have considered in this paper.

\begin{figure}
    \centering
    \resizebox{0.9\columnwidth}{!}{\includegraphics{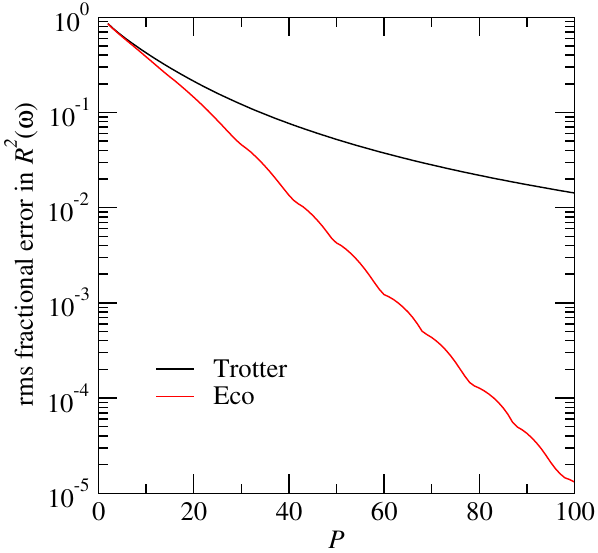}}
    \caption{Root-mean-square fractional errors in $R^2(\omega)$ for the Trotter and Eco path integrals when $\beta\hbar\omega_{\rm max}=50$, as a function of $P$.}
    \label{fig:1}
\end{figure}

Fig.~1 shows the resulting rms fractional error in $R^2(\omega)$ as a function of $P$ for $x_{\rm max}=50$, which is similar to the value we shall use in our ice calculations in Sec.~III ($x_{\rm max}=57.55$). The results for the Trotter path integral are also shown for comparison. The rms fractional error tends to 1 in both cases as $P\to 1$ and the path integral collapses to a single classical bead. As $P$ increases beyond 1, the rms error in the 2nd order Trotter path integral decreases as $P^{-2}$, whereas the rms error in the Eco path integral decreases exponentially (note the log scale on the $y$ axis). Of course this does not imply that the convergence of the Eco path integral will also be exponential when it is applied to anharmonic problems: it simply shows that one can obtain exponential convergence of the {\em harmonic} radius of gyration within the frequency range $0\le\omega\le\omega_{\rm max}$.

\begin{figure}[t]
    \centering
    \resizebox{0.75\columnwidth}{!}{\includegraphics{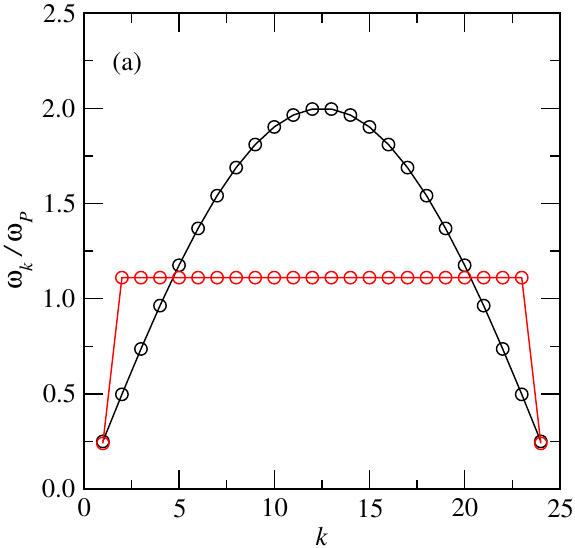}}    
    \vspace{0.5em}
    \resizebox{0.75\columnwidth}{!}{\includegraphics{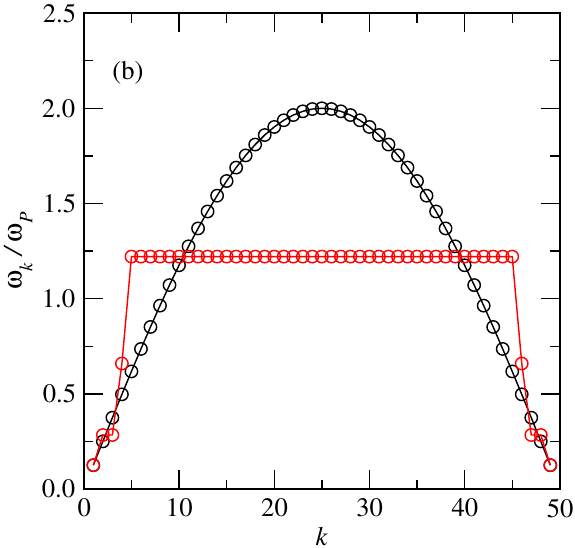}}
    \vspace{0.5em}
    \resizebox{0.75\columnwidth}{!}{\includegraphics{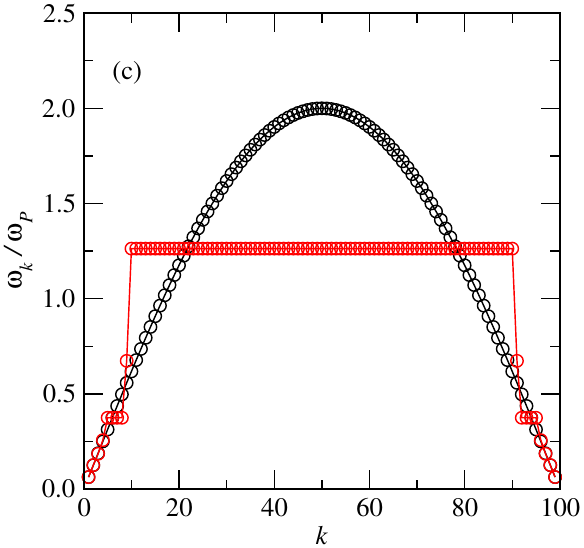}}
    \caption{Comparison of Trotter (black) and Eco (red) normal mode frequencies when $\beta\hbar\omega_{\rm max}=50$ and (a) $P=25$, (b) $P=50$, (c) $P=100$.}
    \label{Fig:2}
\end{figure}

The resulting Eco frequencies are compared with the Trotter frequencies for $P=25$, 50, and 100 in Fig.~2. The optimum fit to the harmonic oscillator model results in a group of degenerate normal modes at a frequency just above $\omega_P$, along with some additional modes at low and high $k$ with frequencies close to those of the Trotter approximation. There is a higher proportion of these low-frequency modes as $P$ increases, suggesting that the Eco frequencies might all eventually tend to the Trotter frequencies in the large $P$ limit where the Trotter path integral is fully converged. However, that limit is clearly far beyond the largest value of $P$ we have considered here, and also far beyond any value of $P$ one would be likely to use in a path integral calculation. A more practical point is that, since the highest Eco normal mode frequency in Fig.~2 is well below the highest Trotter frequency, one might be able to get away with a slightly larger time step in an Eco path integral calculation than in a Trotter calculation. 

\begin{figure}[t]
    \centering
    \resizebox{0.75\columnwidth}{!}{\includegraphics{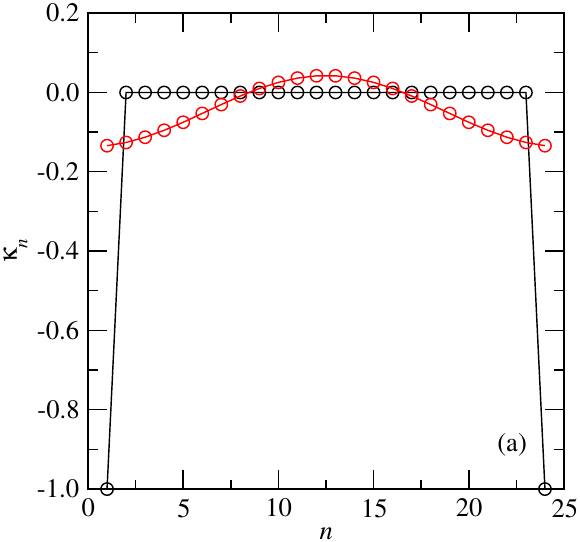}}    
    \vspace{0.5em}
    \resizebox{0.75\columnwidth}{!}{\includegraphics{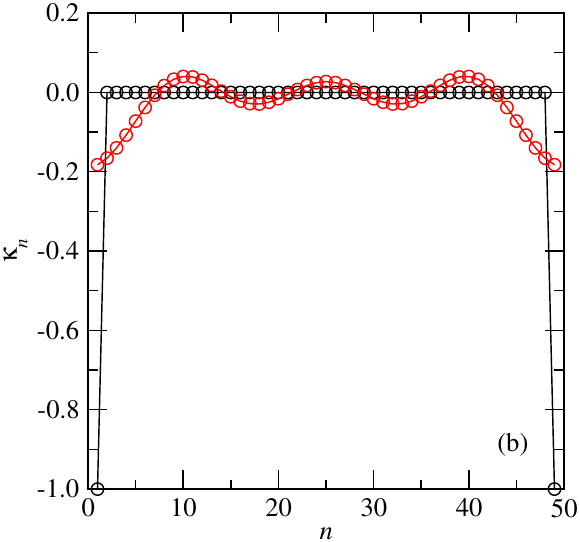}}
    \vspace{0.5em}
    \resizebox{0.75\columnwidth}{!}{\includegraphics{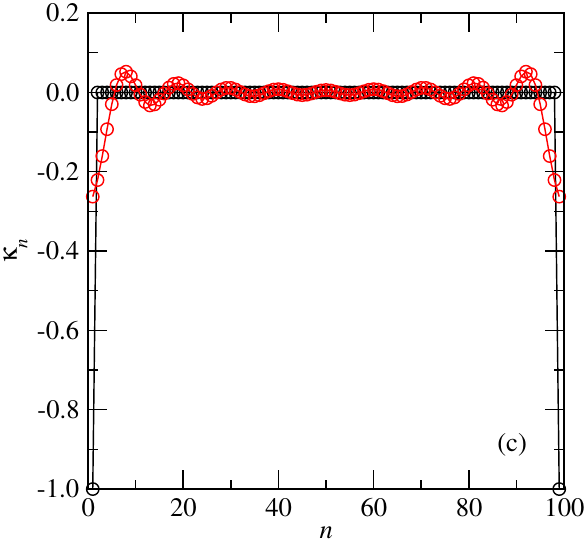}}
    \caption{Comparison of Trotter (black) and Eco (red) circulant coefficients when $\beta\hbar\omega_{\rm max}=50$ and (a) $P=25$, (b) $P=50$, (c) $P=100$. (We have omitted $\kappa_0$ from these plots because it is minus the sum of the other coefficients and to include it in the Trotter case would almost triple  the range of the $y$ axis.)}
    \label{Fig:3}
\end{figure}

Once the Eco frequencies have been found the corresponding circulant coefficients $\kappa_n$ can be calculated using Eq.~(11). Fig.~3 compares these coefficients with those of the Trotter approximation for the three cases considered in Fig.~2. Whereas the Trotter spring potential only contains attractive interactions between nearest neighbors on the ring polymer necklace, the Eco spring potential contains both attractive $(k_n<0)$ and repulsive $(k_n>0)$ interactions between the beads all the way around the ring. The attractive interactions are weaker than in the Trotter case, and for small $P$ they are between close neighbors whereas the repulsive interactions are between distant beads. These features increase $R^2(\omega)$ from its Trotter value, which underestimates the correct result. For larger $P$, an oscillatory pattern develops in the attractive and repulsive coefficients that brings them closer to their Trotter counterparts. However, there is still a significant difference between the Eco and Trotter circulant coefficients even when $P=100$, as was the case for the normal mode frequencies in Fig.~2. 

\subsection{Estimators for observables}

Most other schemes for accelerating the convergence of path integrals, such as the fourth-order Takahashi-Imada\cite{Takahashi1984} and Suzuki-Chin\cite{Suzuki1995,Chin1997} methods, require the derivation of new path integral estimators for observables.\cite{Kapil2016} However, this is not necessary for the Eco path integral because the standard Trotter estimators remain valid for any spring constant matrix $K$ that has the symmetric circulant structure in Eqs.~(5) to~(7).

Consider, for example, the average energy of the system with the partition function $Z_P$ in Eq.~(2),
\begin{equation}
    E = -\frac{\partial \ln Z_P}{\partial \beta} = 
    -\frac{1}{P}\frac{\partial \ln Z_P}{\partial \beta_P},
\end{equation}
when the ring polymer spring potential is defined as ${\sf U}(q)=\frac{1}{2}q^TKq$. Using the fact that $\omega_P=1/(\beta_P\hbar)$, and therefore $\partial (\beta_PK)/\partial \beta_P=-K$ from Eq.~(5), it is straightforward to show that
\begin{equation}
    E = \left< {\cal T}^{\rm TD} \right> + \left< {\cal V}\right>,
\end{equation}
where ${\cal T}^{\rm TD}$ is the usual thermodynamic kinetic energy estimator
\begin{equation}
{\cal T}^{\rm TD} = \frac{P}{2\beta}-\frac{1}{P}{\sf U}(q),
\end{equation}
${\cal V}$ is the usual potential energy estimator 
\begin{equation}
{\cal V} = \frac{1}{P}{\sf V}(q) = \frac{1}{P}\sum_{j=0}^{P-1} V(q_j),
\end{equation}
and the angular brackets denote the configurational average
\begin{equation}
    \left<\cdots\right> = \dfrac{\int {\rm d}q\, e^{-\beta_P\left[{\sf U}(q)+{\sf V}(q)\right]}\left(\cdots\right)}
    {\int {\rm d}q\, e^{-\beta_P\left[{\sf U}(q)+{\sf V}(q)\right]}}.
\end{equation}

The variance of the thermodynamic kinetic energy estimator increases with $P$ because the (one-dimensional) ideal gas term $P/2\beta$ in Eq.~(20) corresponds to a gas at an elevated temperature $P/\beta$. Since this becomes problematic at large $P$, ${\cal T}^{\rm TD}$ is usually replaced with the centroid virial estimator\cite{Herman1982}
\begin{equation}
{\cal T}^{\rm CV} = \frac{1}{2\beta}+\frac{1}{2P}\sum_{j=0}^{P-1} (q_j-\bar{q})\frac{\partial {\sf V}(q)}{\partial q_j}
\end{equation}
in Trotter path integral calculations. That this can also be done in an Eco calculation follows from the identity
\begin{equation}
e^{-\beta_P{\sf V}(q)}\frac{\partial {\sf V}(q)}{\partial q_j}=-\frac{1}{\beta_P}\frac{\partial}{\partial q_j} e^{-\beta_P{\sf V}(q)},
\end{equation}
which can be used to rearrange the expectation value of ${\cal T}^{\rm CV}$ as follows:
\begin{align*}
    &\int {\rm d}q\,e^{-\beta_P[{\sf U}(q)+{\sf V}(q)]} (q_j-\bar{q})\frac{\partial {\sf V}(q)}{\partial q_j}\\
    &=-\frac{1}{\beta_P}\int {\rm d}q\, e^{-\beta_P{\sf U}(q)}(q_j-\bar{q})\frac{\partial}{\partial q_j}e^{-\beta_P{\sf V}(q)}\\
    &=+\frac{1}{\beta_P}\int {\rm d}q\, e^{-\beta_P{\sf V}(q)}\frac{\partial}{\partial q_j}\left[(q_j-\bar{q})e^{-\beta_P{\sf U}(q)}\right]\\
    &=\frac{1}{\beta_P}\int {\rm d}q\, e^{-\beta_P[{\sf U}(q)+{\sf V}(q)]}\left[1-\frac{1}{P}-\beta_P(q_j-\bar{q})\frac{\partial{\sf U}(q)}{\partial q_j}\right].
\end{align*}
Since $\partial {\sf U}(q)/\partial q_j = \sum_{i=0}^{P-1} K_{ji}q_i$ from Eq.~(4), and $\sum_{j=0}^{P-1} K_{ji}=0$ from Eqs.~(5) and~(7), this gives
\begin{align}
\left<{\cal T}^{\rm CV}\right> &= \left<\frac{1}{2\beta_P}-\frac{1}{2P}q^TKq\right>\nonumber\\
&=\left<\frac{P}{2\beta}-\frac{1}{P}{\sf U}(q)\right>\nonumber\\
&=\left<{\cal T}^{\rm TD}\right>.
\end{align}
The ensemble average of ${\cal T}^{\rm CV}$ is thus the same as that of ${\cal T}^{\rm TD}$, but since its ideal gas term is at the correct temperature its fluctuations are significantly smaller.\cite{Herman1982}

Another way to derive the potential energy estimator in Eq.~(21) is to note that, since $V(\hat{q})|q_0\rangle=|q_0\rangle V(q_0)$, the path integral approximation to
\begin{equation}
V = \frac{1}{Z}{\rm tr}\left[e^{-\beta\hat{H}}V(\hat{q})\right]
\end{equation}
is simply $\left<V(q_0)\right>$. Provided $H_P(p,q)$ is invariant to a cyclic permutation of the beads one can then re-write this as $\frac{1}{P}\sum_{j=0}^{P-1} \left<V(q_j)\right>$. Since this argument works equally well in the Eco case as the Trotter case, both for $V(\hat{q})$ and for any other local (coordinate-dependent) operator $A(\hat{q})$, we have
\begin{equation}
{\cal A} = \frac{1}{P}\sum_{j=0}^{P-1} A(q_j)
\end{equation}
more generally. Hence a radial distribution function (for example) can be calculated from a histogram of ring polymer bead distances in an Eco calculation in the same way as in a Trotter calculation.

\section{Application to hexagonal ice}

As a first test of the Eco path integral in a more realistic (anharmonic) setting, we shall now present some results for hexagonal ice. For these calculations, we used the q-TIP4P/F water potential\cite{Habershon2009} to study the proton-disordered $3\times 2\times 2$ unit cell of Hayward and Reimers,\cite{Hayward1997} which contains $N=96$ water molecules with no net dipole moment. Our PIMD calculations were performed at 100 K and the experimental density (0.993 g cm$^{-3}$), with a global velocity rescaling thermostat\cite{Bussi2007} on the centroid and a local PILE thermostat\cite{Ceriotti2010} on each internal normal mode. The Eco frequencies were calculated using $\omega_{\rm max}=4000$ cm$^{-1}$, which gives $x_{\rm max}=57.55$ at 100 K. Everything else in the Eco calculation was done with a standard Trotter path integral code.

\subsection{Kinetic and potential energies}

\begin{figure}[t]
    \centering
    \resizebox{0.75\columnwidth}{!}{\includegraphics{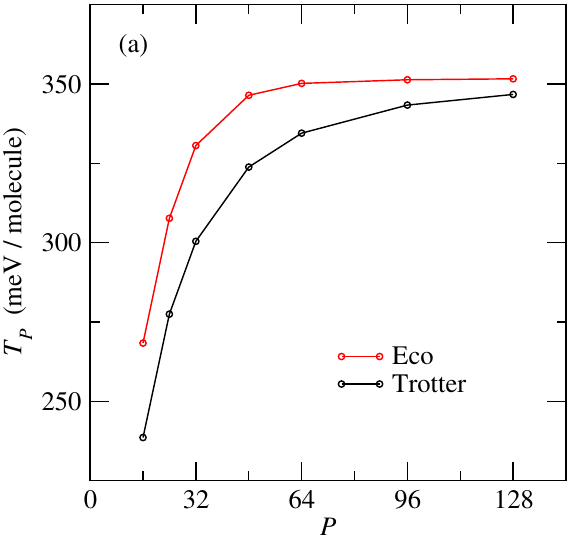}}    
    \vspace{0.5em}
    \resizebox{0.75\columnwidth}{!}{\includegraphics{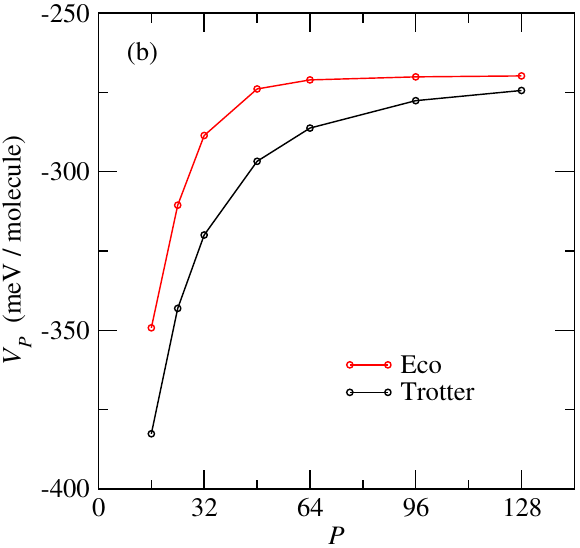}}
    \caption{Comparison of Trotter and Eco (a) kinetic and (b) potential energies of hexagonal ice at 100 K, as a function of the number of path integral beads.}
    \label{Fig:4}
\end{figure}

Fig.~4 shows how the kinetic and potential energies obtained from the Trotter and Eco calculations converge towards the exact result with increasing $P$. The Eco path integral clearly converges significantly faster than the Trotter path integral for this anharmonic problem. The kinetic and potential energies obtained from the Eco calculation with $P=48$ are better converged than those obtained from the Trotter calculation with $P=128$. The convergence of $T_P$ and $V_P$ is monotonic in the Trotter calculation and it remains monotonic in the Eco calculation, which makes it straightforward to see when to stop increasing $P$.

\begin{figure}[t]
    \centering
    \resizebox{0.8\columnwidth}{!}{\includegraphics{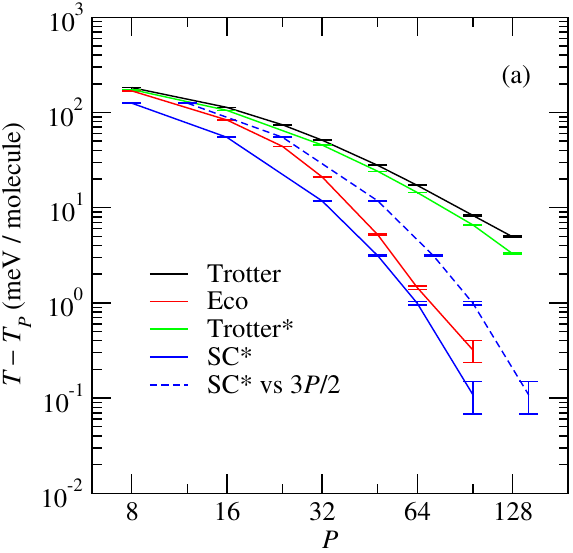}}    
    \vspace{0.5em}
    \resizebox{0.8\columnwidth}{!}{\includegraphics{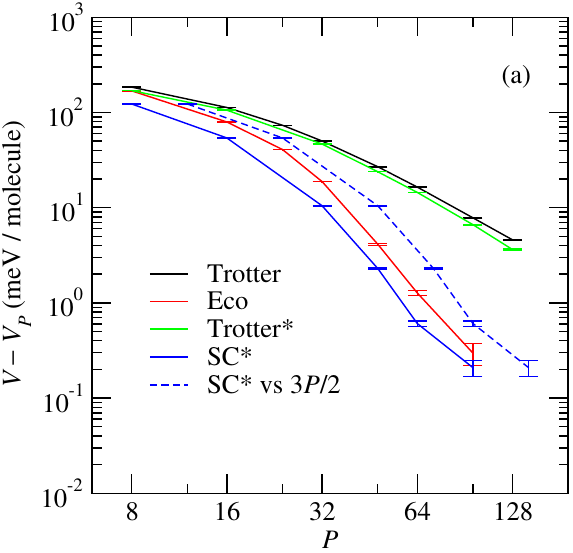}}
    \caption{Errors in the Trotter and Eco (a) kinetic and (b) potential energies of hexagonal ice at 100 K as a function of $P$, plotted on a log-log scale. The starred Trotter and 4th order SC results from Ref.~\cite{Kapil2016} are also shown for comparison, with the latter plotted as a function of both $P$ and the number of force evaluations in the projected Hessian SC method, $3P/2$.}
    \label{Fig:5}
\end{figure}

Since the differences between the Eco results for $P=96$ and $P=128$ are already well below a millivolt per molecule, we shall stop there and regard the Eco $P=128$ calculation as well converged. This provides a way to plot the errors in $T_P$ and $V_P$ from the two methods as a function of $P$, which we have done on a log-log scale in Fig.~5. (Here the error bars are standard errors in the mean from 8 independent 10 ps PIMD trajectories, with the standard error in the $P=128$ Eco reference calculation added to the error of each data point.) The large $P$ slopes of the Trotter curves in this figure are just above $-2$, as expected for a 2nd order method, whereas the large $P$ slopes of the Eco curves are just above $-4$. Thus the exponential convergence of the harmonic radius of gyration (and therefore also the harmonic quantum energy shift) of the Eco path integral  in Fig.~1 does not carry over to this anharmonic problem, where the convergence is more like that of a 4th order method. 

Kapil, Behler, and Ceriotti\cite{Kapil2016} have presented a very similar plot in their paper on higher-order path integrals. Although they used a different (machine-learned) potential in their study, the Trotter results in their Fig.~3 are almost identical to our Trotter results, as can be seen from the direct comparison we have included in Fig.~5. This legitimizes a comparison between their 4th order Suzuki-Chin (SC) results and our Eco results, which are again very similar. While the SC path integral is slightly more accurate than the Eco path integral for a given value of $P$, this is not a like-for-like comparison, because Kapil {\em et al.} used an extra force evaluation to evaluate the projected Hessian that arises on each even bead in the SC method by finite differences.\cite{Kapil2016} To take this extra effort into account, one should compare a $P$-bead SC calculation with a $3P/2$-bead Eco calculation, both of which involve the same number of force evaluations. When this is done we find that the Eco calculation has a higher accuracy per force evaluation than the SC method throughout the range considered in Fig.~5. Note also that Kapil {\em et al.} had to develop new path integral estimators for their projected Hessian SC method,\cite{Kapil2016} which we have not needed to do for the Eco path integral for the reasons explained in Sec.~II.C. Thus the Eco method is both simpler and more efficient than projected Hessian SC.\cite{SCeffort}

\subsection{Radial distribution functions}

\begin{figure}[t]
    \centering
    \resizebox{0.9\columnwidth}{!}{\includegraphics{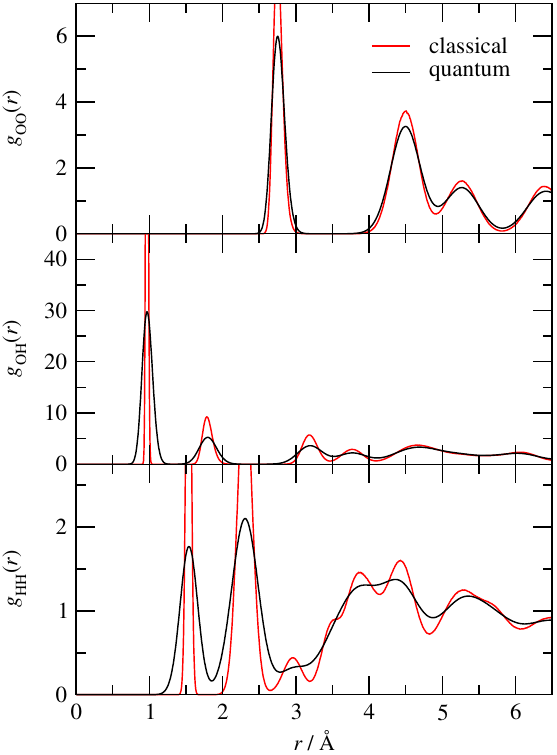}}    
    \caption{Quantum and classical radial distribution functions of q-TIP4P/F ice Ih at 100 K.}
    \label{Fig:6}
\end{figure}

Fig.~6 compares the converged quantum and classical radial distribution functions of q-TIP4P/F hexagonal ice at 100 K. The quantum (bead) distribution functions were obtained from an Eco PIMD reference simulation with $P=128$, and the classical distribution functions were obtained using the same code with $P=1$. One sees that while the classical O--O distribution function exhibits the intermediate-range order expected for a crystalline oxygen sublattice, this order is washed out in the O--H and H--H distribution functions by both the proton disorder and the greater quantum delocalisation of the hydrogen atoms.

\begin{figure}[t]
    \centering
    \resizebox{0.8\columnwidth}{!}{\includegraphics{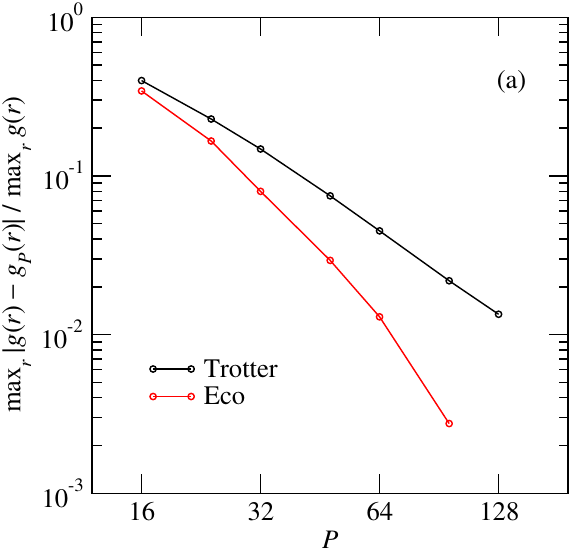}}    
    \vspace{0.5em}
    \resizebox{0.8\columnwidth}{!}{\includegraphics{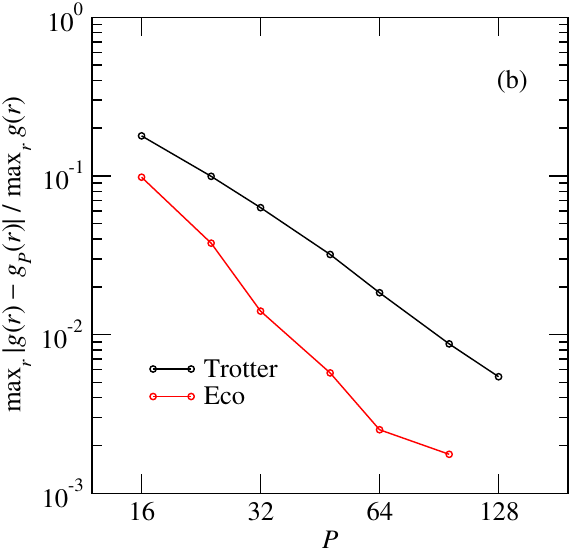}}
    \caption{Fractional errors in the Trotter and Eco (a) O--H and (b) H--H radial distribution functions as a function of $P$, on a log-log scale.}
    \label{Fig:7}
\end{figure}

The features in Fig.~6 that are most difficult to converge in a PIMD simulation are the intra-molecular O--H and H--H peaks, which are significantly narrower in the classical calculation than the quantum calculation. Fig.~7 shows how these peaks converge with increasing $P$ by plotting the fractional errors $\max_r |g(r)-g_P(r)|/\max_r g(r)$ in the Trotter and Eco radial distribution functions relative to the 128-bead Eco reference calculation. Since one is usually only interested in calculating radial distribution functions to ``graphical" ($\sim 1\%$) accuracy, what matters most in this figure is where the curves pass through a fractional error of around $10^{-2}$. For the O--H distribution function, this happens just beyond $P=64$ for the Eco path integral and just beyond $P=128$ for the Trotter path integral. Both methods converge more rapidly for the H--H distribution function, presumably because the water bending frequency is lower than the O--H stretching frequency.  The upshot is thus that convergence of all three radial distribution functions to graphical accuracy requires around half as many beads in the Eco calculation, and if higher accuracy were required the Eco speedup would be even greater.

\subsection{Vibrational density of states}

In their paper on higher-order path integrals, Kapil {\em et al.}\cite{Kapil2016} compared projected Hessian SC and Trotter calculations of the vibrational density of states (VDOS) of liquid water at 300 K using the thermostatted ring polymer molecular dynamics (TRPMD) method.\cite{Rossi2014} They found that, while both eventually converged to the same result, the convergence of the 4th order SC path integral was {\em slower} than that of the 2nd order Trotter path integral because of a spurious shift in physical vibrational frequencies caused by the projected Hessian SC force.\cite{Kapil2016} 

We shall not consider TRPMD here because there are now better path integral methods for vibrational spectra. Quasi-centroid molecular dynamics (QCMD),\cite{Trenins2019} fast QCMD,\cite{Fletcher2021} and the elevated temperature path integral ground state (T$_e$ PIGS) method\cite{Musil2022} have all been shown to eliminate the problems of previous approaches (including TRPMD) and to give good agreement with one another for the spectra of water and ice.\cite{Lawrence2023} We have therefore selected the simplest of these methods (T$_e$ PIGS) for our first example application of the Eco path integral to a dynamical problem and used it to calculate the VDOS of q-TIP4P/F ice.

\begin{figure}[t]
    \centering
    \resizebox{0.9\columnwidth}{!}{\includegraphics{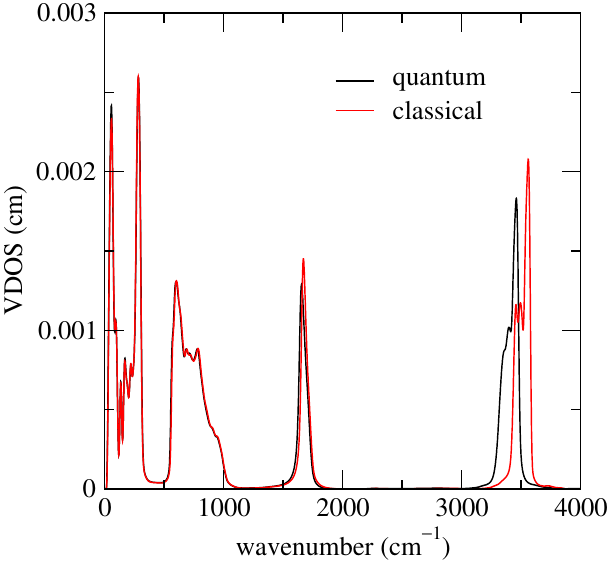}}    
    \caption{Classical ($P=1$) and Eco ($P=32$) vibrational densities of states of q-TIP4P/F ice Ih at 100 K.}
\end{figure}

Rather than machine-learn the elevated temperature centroid potential of mean force in the T$_e$ PIGS method, we have found it simpler to use an approximate on-the-fly implementation based on the adiabatic CMD algorithm.\cite{Cao1996} This implementation is described in Appendix~A. It involves assigning artificial masses to the internal modes of the ring polymer so that they oscillate at a common frequency $\Omega\gg \omega_{\rm max}$, attaching a thermostat to these internal modes to sample them canonically at the elevated temperature $T_e$, and reducing the PIMD integration time step accordingly. In our calculations we used an adiabatic separation frequency of $\Omega=40,000$ cm$^{-1}$ and a time step of 0.01 fs to evolve the internal modes. We also followed Musil {\em et al.}\cite{Musil2022} in setting the elevated temperature to 600 K, which is low enough to put the intramolecular vibrations in their ground quantum states and yet high enough to avoid any issues associated with the CMD curvature problem.\cite{Ivanov2010} The physical temperature was set to 100 K.

\begin{figure}[t]
    \centering
    \resizebox{0.8\columnwidth}{!}{\includegraphics{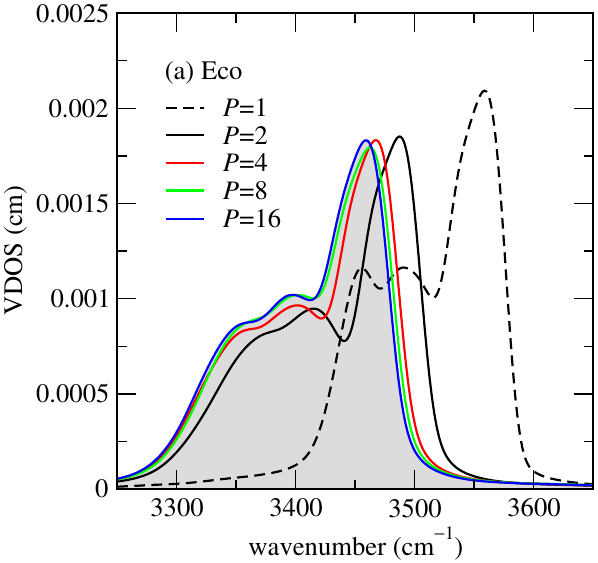}}    
    \vspace{0.5em}
    \resizebox{0.8\columnwidth}{!}{\includegraphics{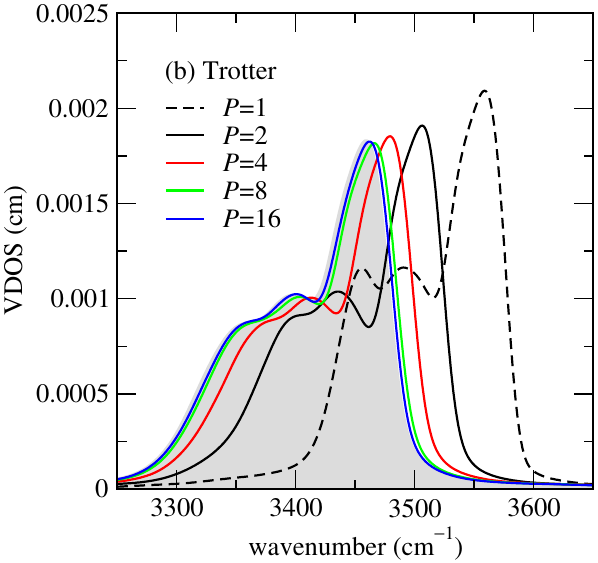}}
    \caption{Convergence of (a) Eco and (b) Trotter T$_{\rm e}$ PIGS calculations of the ice VDOS in the O--H stretching region with respect to $P$. The shaded gray area is from a fully converged Eco calculation with $P=32$ and the dashed line is the classical VDOS with $P=1$.}
\end{figure}

Fig.~8 compares the VDOS of ice at this temperature obtained from a fully converged ($P=32$) Eco calculation with the purely classical ($P=1$) result. One sees that the predicted quantum effects in the spectrum are comparatively minor, in agreement with previous studies.\cite{Trenins2019,Musil2022,Lawrence2023} There are no noticeable quantum effects on the oxygen sublattice phonons below 500 cm$^{-1}$, or on the librational phonons below 1000 cm$^{-1}$. The only noticeable effects are a slight red-shifting and broadening of the intramolecular bending band at $\sim 1650$ cm$^{-1}$, and a more significant red-shifting and broadening of the intramolecular stretching band at $\sim 3500$ cm$^{-1}$, both as a result of anharmonicity. 

Fig.~9 focuses on the larger of these effects -- the red-shifting and broadening of the intramolecular stretching band -- and shows how it converges with increasing $P$ in Eco and Trotter calculations. In contrast to what Kapil {\em et al.}\cite{Kapil2016} found when comparing the convergence of their Trotter and Suzuki-Chin TRPMD calculations for the VDOS of liquid water, we find here that the Eco calculation converges slightly {\em faster} than the Trotter calculation. This is not especially important because the elevated temperature aspect of T$_{\rm e}$ PIGS makes both path integrals easy to converge, but it does at least show that there is no reason not to use the Eco path integral in preference to the Trotter path integral for this sort of dynamical calculation.

\section{Application to MOF-5}

The above calculations on hexagonal ice were performed with an in-house computer program designed for use with empirical potentials such as the q-TIP4P/F water model. To make the Eco path integral more broadly applicable and widely accessible, we have also written a patch to include it in the GPUMD software package,\cite{Fan2022} which provides an extremely efficient way to do PIMD simulations on GPUs \cite{Ying2025,Zeng2025} and which can both train and run machine-learned neuroevolution potentials (NEPs).\cite{Fan2021} Our patch is available on GitHub.\cite{Zeng2026}

To provide an illustration of the functionality of the patch, we first used vanilla GPUMD to train a machine-learned neuroevolution potential for metal--organic framework 5 (MOF-5) using the DFT reference data generated by Wieser and Zojer.~\cite{Wieser2024} We then used the patched version to run Eco and Trotter PIMD calculations on a 2$\times$2$\times$2 supercell containing 3328 atoms constructed from the orthorhombic 416-atom MOF-5 unit cell shown in Fig.~10. The Eco optimisation was done with a maximum frequency of $\omega_{\rm max}=3500$ cm$^{-1}$ obtained from a preliminary calculation of the harmonic phonon dispersion curves. MOF-5 was chosen as an illustration because it has some unusual properties that are difficult to converge using a Trotter path integral, such as its negative thermal expansion coefficient\cite{Lock2010} and a subtle temperature-dependence of its constant pressure heat capacity\cite{Kloutse2015} at low temperatures. We shall now discuss each of these properties in turn.

\begin{figure}[t]
   \centering
   \resizebox{0.9\columnwidth}{!}{\includegraphics{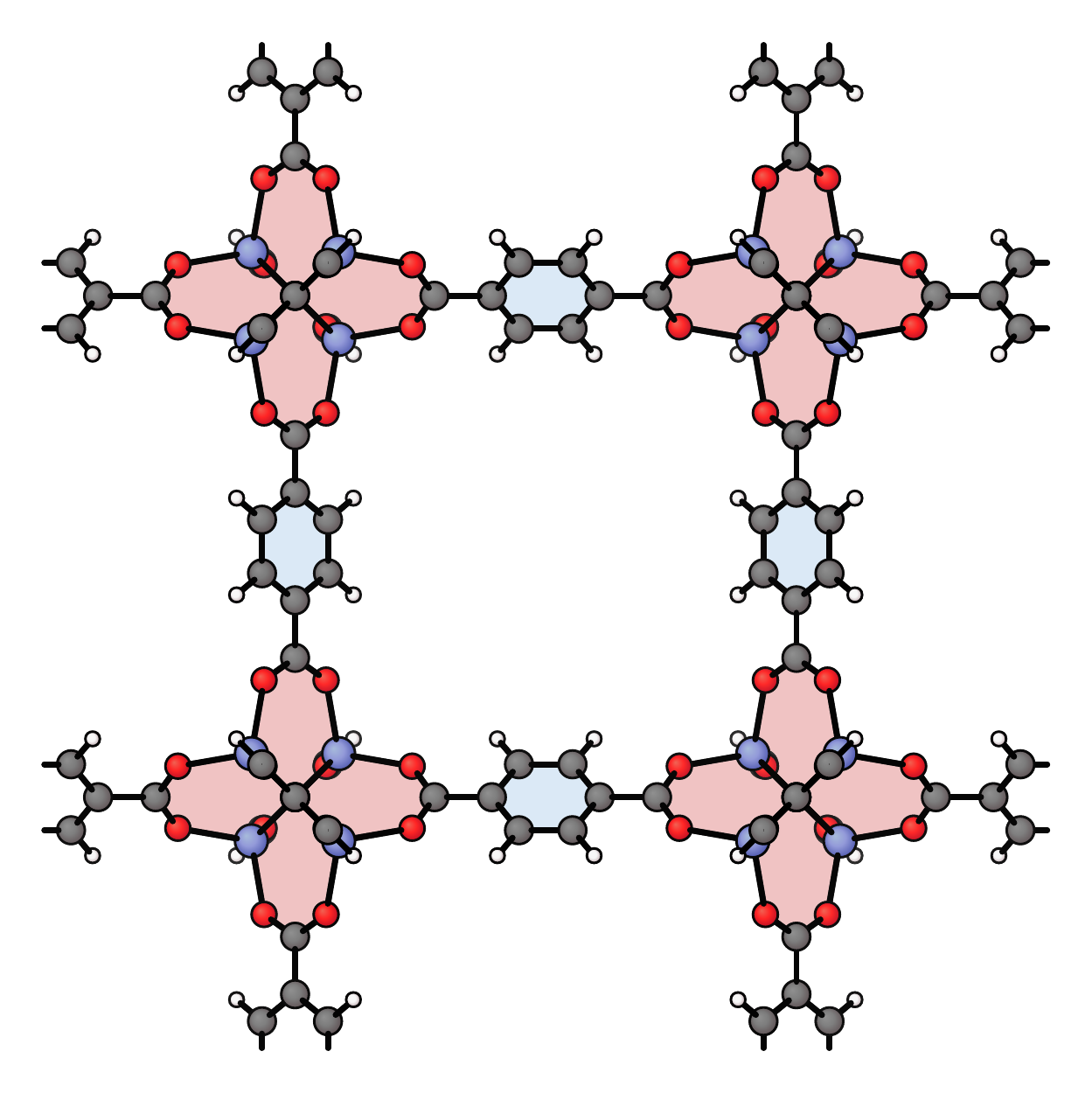}}    
   \caption{Two-dimensional projection of the MOF-5 crystal structure, showing the connectivity between inorganic Zn--O clusters and organic linker units. Grey, white, red, and blue spheres denote C, H, O, and Zn atoms, respectively. This plot was generated using the \texttt{xyzrender} package.\cite{Goodfellow2026} }
\end{figure}

\subsection{Negative thermal expansion coefficient}

Fig.~11 compares the temperature dependence of the Eco and Trotter lattice constants of MOF-5, $\bar a=(a+b+c)/3$, obtained from zero-pressure NPT PIMD simulations. 
The dashed grey lines show the experimental slope,\cite{Lock2010} which we have shifted vertically at regular intervals to facilitate a comparison with the simulations. Both Eco and Trotter recover a negative slope at sufficiently large bead number, but their convergence with respect to $P$ is markedly different.
For Eco PIMD, the results for $P=32$, 64, and 128 are already very close to one another and give a slope consistent with the experimental negative thermal expansion coefficient. By contrast, the Trotter path integral converges more slowly: the small-$P$ Trotter results show larger changes in both the absolute value of $\bar a$ and its temperature dependence, and somewhere between 64 and 128 beads are required to reach the level of convergence  obtained with Eco PIMD at $P=32$. These results are fully consistent with those we have reported above for hexagonal ice, but now in a significantly more complicated setting.

\begin{figure}[t]
    \centering
   \resizebox{0.9\columnwidth}{!}{\includegraphics{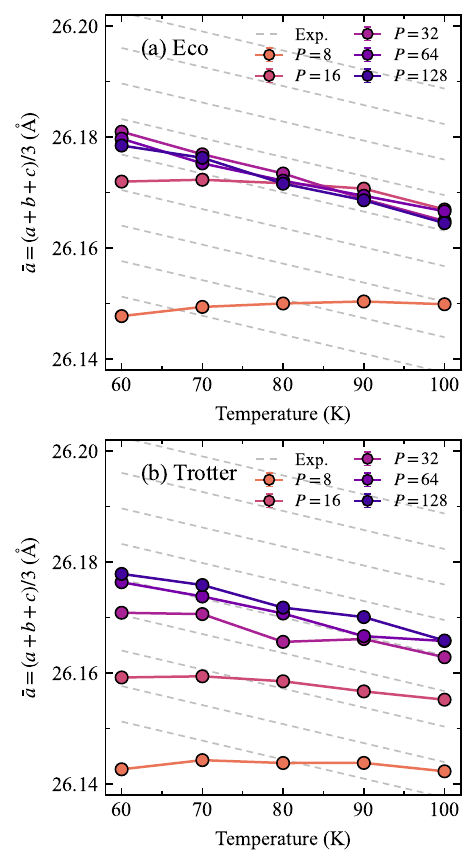}}    
   \caption{Temperature dependence of the average lattice constant, $\bar{a}=(a+b+c)/3$, of MOF-5 calculated using (a) Eco and (b) Trotter PIMD. The slope of the grey dashed lines indicates the experimental linear thermal expansion coefficient,\cite{Lock2010} ${\rm d}\ln\bar{a}/{\rm d}T=-13.1\times10^{-6}\,\mathrm{K}^{-1}$.}
\end{figure}

\subsection{Temperature dependence of the heat capacity}

Our final example application of the Eco path integral is to the temperature dependence of the constant pressure heat capacity of MOF-5. Since this depends on the {\em second} derivative of the energy with respect to temperature, it is even more difficult to converge than the heat capacity itself, which is already a notoriously difficult problem for path integral methods because of its sensitivity to high frequency modes.\cite{Markland2018}

Rather than use a path integral estimator for the heat capacity,\cite{Yamamoto2005} we prefer to use a two-step procedure that only requires the centroid virial NVT energy estimator. At each temperature on a suitably spaced grid, we first perform an NPT PIMD simulation at zero external pressure to determine the corresponding equilibrium lattice constant $\bar a$. We then fix this lattice constant and perform a NVT PIMD simulation to obtain the total energy $E$ (or equivalently the enthalpy as the external pressure is zero). We fit the resulting $E(T)$ to a quadratic function, taking proper account of its statistical error at each temperature, and obtain $C_p(T)$ from the temperature derivative of the fitted curve. 

Fig.~12 shows the convergence of the total energy and the resulting heat capacity of MOF-5 for Eco and Trotter PIMD. The slope of $E(T)$ in the Trotter results for $P=64$ and $P=128$ is clearly too large, leading to a substantial overestimation of the heat capacity. Moreover the curvature of the Trotter $E(T)$ is still slightly negative even when $P=256$, resulting in a heat capacity that decreases with increasing $T$. The Eco calculation converges significantly more rapidly, giving a heat capacity that is already increasing with $T$ when $P=128$ and is in excellent agreement with the experimental measurement\cite{Kloutse2015} when $P=256$. To reach this level of convergence with Trotter PIMD would require at least a thousand beads. This is a more stringent thermodynamic validation of the Eco path integral than the lattice constant alone: Eco PIMD not only captures the negative thermal expansion of MOF-5, it provides a more rapidly convergent estimate of the low-temperature heat capacity while using the same energy estimator as Trotter PIMD.

\begin{figure}[t]
    \centering
    \label{fig:mof5_cp}
   \resizebox{0.9\columnwidth}{!}{\includegraphics{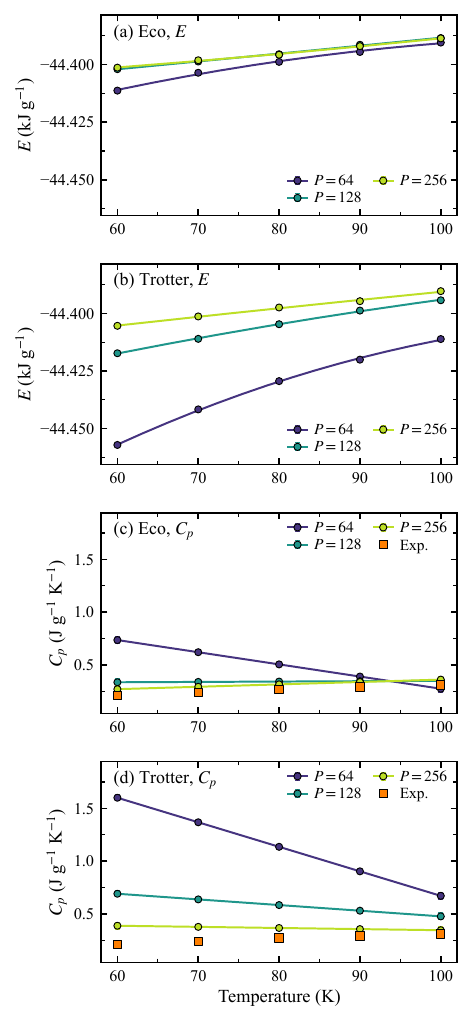}}    
   \caption{Temperature dependence of the total energy, $E$, and heat capacity, $C_p$, of MOF-5 calculated using Eco and Trotter PIMD. Panels (a) and (b) show $E(T)$, while panels (c) and (d) show $C_p(T)$, for different bead numbers $P$. Orange squares denote the experimental\cite{Kloutse2015} heat capacity.}
\end{figure}

\section{Conclusion}

We have developed an economised path integral and shown that it can be used to calculate both static equilibrium properties with PIMD and dynamical properties with methods like T$_{\rm e}$ PIGS.\cite{Musil2022,Castro2025} The Eco path integral outperforms the Trotter path integral in every case we have considered, offering a modest improvement for low accuracy work and a substantial improvement at higher accuracy. The reduction in $P$ relative to Trotter is comparable to that of the 4th order projected Hessian SC method,\cite{Kapil2016} but comes with purely 2nd order Trotter effort. There is no need to derive any new estimators, no need to calculate any projected Hessians, and very little else that needs to be changed in a standard Trotter path integral code. \rev{Since these advantages suggest that Eco may soon become the default method for applications like those in this paper, we have provided both a reference version of our Eco optimisation code in the supplementary material and an Eco patch to GPUMD on GitHub\cite{Zeng2026} to facilitate its adoption.} 

\section*{Supplementary Material}

Section~A of the supplementary material contains a listing of the Fortran code we used to produce Figs.~1-3. Section~B compares the ``pigs might fly" method described in the Appendix  with both PA-Te-CMD and fully adiabatic Te PIGS calculations of the dipole absorption spectrum of ice at 150 K. 

\begin{acknowledgments}
We would like to thank Jorge Castro, George Trenins, and Mariana Rossi for an interesting discussion about the ``pigs might fly" method described in the Appendix and its relationship to their PA-Te-CMD method.\cite{Castro2025} Z.Z. acknowledges support from a Newton International Fellowship from The Royal Society (No.~NIF/R1/254124) and an allocation of computer time on the Oxford ARC system.
\end{acknowledgments}

\section*{Data Availability}

The data that support the findings of this study are available within the article and the supplementary material. 
\appendix

\section{Pigs might fly}

The T$_{\rm e}$ PIGS method is classical molecular dynamics at the target temperature $T$ on a centroid potential of mean force generated at an elevated temperature $T_e$.\cite{Musil2022} In this Appendix we use the one-dimensional notation we have been using throughout this paper to describe an approximate ``on-the-fly" implementation of this method based on the adiabatic CMD algorithm.\cite{Cao1996} 

The goal of our implementation is to keep the dynamics at the reciprocal temperature $\beta=1/k_{\rm B}T$ while ensuring that the fluctuations of the internal mode coordinates are at the reciprocal temperature $\beta_e=1/k_{\rm B}T_e$. This can be accomplished by re-writing Eq.~(2) as
\begin{equation}
    H_P(p,q) = H_P^{0}(p,q)+\sum_{j=0}^{P-1}V(q_j),
\end{equation}
and replacing the usual free ring polymer Hamiltonian $H_P^{(0)}(p,q)$ with
\begin{equation}
\tilde{H}_P^{(0)}(p,q) = \frac{\tilde{p}_0^2}{2m}+\sum_{k=1}^{P-1} \left[\frac{\tilde{p}_k^2}{2m_k}+\frac{1}{2}m\tilde{\omega}_{k}^2\tilde{q}_k^2\right],
\end{equation}
where $\tilde{\omega}_k=\sqrt{{\beta_e}/{\beta}}\,\omega_{k,e}$, $\omega_{k,e}$ is a (Trotter or Eco) internal mode frequency at the elevated temperature $T_e$, and
\begin{equation}
\tilde{p}_k = \sum_{j=0}^{P-1} p_jC_{jk}\quad\hbox{and}\quad \tilde{q}_k = \sum_{j=0}^{P-1} q_jC_{jk}
\end{equation}
are the normal mode momenta and coordinates. The modified Boltzmann density
\begin{equation}
    \tilde{\rho}_P(p,q) \propto e^{-\beta_P\left[\tilde{H}_P^{(0)}(p,q)+\sum_{j=0}^{P-1}V(q_j)\right]}
\end{equation}
of a PIMD simulation at temperature $T$ will then keep the centroid $(k=0)$ mode at this temperature while the fluctuations of the internal ($k>0$) mode coordinates are at the elevated temperature $T_e$.

To implement the adiabatic CMD algorithm\cite{Cao1996} in this context, we choose the internal mode masses $m_k$ in Eq.~(A2) to be such that all $P-1$ internal modes oscillate at the same adiabatically separated frequency $\Omega\gg \omega_{\rm max}$ in the absence of the physical potential:
\begin{equation}
m_k = \left(\frac{\tilde{\omega}_{k}}{\Omega}\right)^2 m.
\end{equation}
We modify the free ring polymer evolution stage of the PIMD algorithm to be consistent with Eq.~(A2), and we attach a local PILE thermostat\cite{Ceriotti2010} with a friction coefficient of $\gamma=\Omega$ and a target temperature of $T$ to each $\tilde{p}_k$ to ensure that the fluctuations of $\tilde{q}_k$ are at temperature $T_e$. A global velocity rescaling thermostat\cite{Bussi2007} is applied to the centroid during an initial equilibration stage, but then switched off for the centroid dynamics calculation. 

\rev{We should point out that Castro {\em et al.}\cite{Castro2025} have proposed an alternative to this approach that they describe as partially adiabatic elevated temperature centroid molecular dynamics (PA-Te-CMD). Their method uses two separate thermostats to keep the centroid at temperature $T$ and the internal modes at temperature $T_e$ in a non-equilibrium steady-state calculation, whereas our method involves a thermal equilibrium simulation at the physical temperature $T$ with a modified Hamiltonian that keeps the fluctuations of the internal mode coordinates at the elevated temperature $T_e$. We have compared the two approaches for the dipole absorption spectrum of ice at 150 K and found that our implementation gives somewhat better agreement with the fully adiabatic Te PIGS spectrum reported by Musil {\em et al.}\cite{Musil2022} (see Sec.~B of the SI). Presumably this is because the two thermostats in the non-equilibrium PA-Te-CMD method struggle to suppress the energy flow from the internal modes to the centroid when $T_e\gg T$.\cite{Castro2025} Since this problem is expected to become even more severe at $T=100$ K, we decided to use our method for the ice VDOS calculations reported in Sec.~III.C.}

\bibliography{ecorefs.bib}

@article{Behler2007,
  author    = {J{\"o}rg Behler and Michele Parrinello},
  title     = {Generalized neural-network representation of high-dimensional potential-energy surfaces},
  journal   = {Phys. Rev. Lett.},
  year      = {2007},
  volume    = {98},
  number    = {14},
  pages     = {146401},
  doi       = {10.1103/PhysRevLett.98.146401},
  url       = {https://doi.org/10.1103/PhysRevLett.98.146401},
  publisher = {American Physical Society}
}

@article{Bartok2010,
  author    = {Albert P. Bart{\'o}k and Mike C. Payne and Risi Kondor and G{\'a}bor Cs{\'a}nyi},
  title     = {Gaussian approximation potentials: The accuracy of quantum mechanics, without the electrons},
  journal   = {Phys. Rev. Lett.},
  year      = {2010},
  volume    = {104},
  number    = {13},
  pages     = {136403},
  doi       = {10.1103/PhysRevLett.104.136403},
  url       = {https://doi.org/10.1103/PhysRevLett.104.136403},
  publisher = {American Physical Society}
}

@article{Chandler1981,
  author    = {David Chandler and Peter G. Wolynes},
  title     = {Exploiting the isomorphism between quantum theory and classical statistical mechanics of polyatomic fluids},
  journal   = {J. Chem. Phys.},
  year      = {1981},
  volume    = {74},
  number    = {7},
  pages     = {4078--4095},
  doi       = {10.1063/1.441588},
  url       = {https://doi.org/10.1063/1.441588},
  publisher = {AIP Publishing}
}

@article{Markland2018,
  author    = {Thomas E. Markland and Michele Ceriotti},
  title     = {Nuclear quantum effects enter the mainstream},
  journal   = {Nat. Rev. Chem.},
  year      = {2018},
  volume    = {2},
  pages     = {0109},
  doi       = {10.1038/s41570-017-0109},
  url       = {https://doi.org/10.1038/s41570-017-0109},
  publisher = {Springer Nature}
}

@article{Hunt2026,
  author  = {Hunt, Andrew C. and Althorpe, Stuart C.},
  title   = {Exploiting the path-integral radius of gyration in open quantum dynamics},
  journal = {J. Chem. Phys.},
  year    = {2026},
  month   = feb,
  volume  = {164},
  number  = {8},
  pages   = {084113},
  doi     = {10.1063/5.0314385},
  url     = {https://doi.org/10.1063/5.0314385},
  publisher = {AIP Publishing}
}

@book{Feynman1965,
  author    = {Richard P. Feynman and Albert R. Hibbs},
  title     = {{Quantum Mechanics and Path Integrals}},
  publisher = {McGraw-Hill},
  address   = {New York},
  year      = {1965}
}

@article{Rossi2014,
  author    = {Mariana Rossi and Michele Ceriotti and David E. Manolopoulos},
  title     = {How to remove the spurious resonances from ring polymer molecular dynamics},
  journal   = {J. Chem. Phys.},
  year      = {2014},
  volume    = {140},
  number    = {23},
  pages     = {234116},
  doi       = {10.1063/1.4883861},
  url       = {https://doi.org/10.1063/1.4883861},
  publisher = {AIP Publishing}
}

@book{Davis1994,
  author    = {Philip J. Davis},
  title     = {{Circulant Matrices}},
  edition   = {2},
  publisher = {AMS Chelsea Publishing},
  address   = {Providence, RI},
  year      = {1994},
  isbn      = {9780821891650}
}

@article{Ceriotti2010,
  author    = {Michele Ceriotti and Michele Parrinello and Thomas E. Markland and David E. Manolopoulos},
  title     = {Efficient stochastic thermostatting of path integral molecular dynamics},
  journal   = {J. Chem. Phys.},
  year      = {2010},
  volume    = {133},
  number    = {12},
  pages     = {124104},
  doi       = {10.1063/1.3489925},
  url       = {https://doi.org/10.1063/1.3489925},
  publisher = {AIP Publishing}
}

@article{Ceriotti2011,
  author    = {Michele Ceriotti and David E. Manolopoulos and Michele Parrinello},
  title     = {Accelerating the convergence of path integral dynamics with a generalized Langevin equation},
  journal   = {J. Chem. Phys.},
  year      = {2011},
  volume    = {134},
  number    = {8},
  pages     = {084104},
  doi       = {10.1063/1.3556661},
  url       = {https://doi.org/10.1063/1.3556661},
  publisher = {AIP Publishing}
}

@article{Ceriotti2012,
  author    = {Michele Ceriotti and David E. Manolopoulos},
  title     = {Efficient first-principles calculation of the quantum kinetic energy and momentum distribution of nuclei},
  journal   = {Phys. Rev. Lett.},
  year      = {2012},
  volume    = {109},
  number    = {10},
  pages     = {100604},
  doi       = {10.1103/PhysRevLett.109.100604},
  url       = {https://doi.org/10.1103/PhysRevLett.109.100604},
  publisher = {American Physical Society}
}

@article{Poltavsky2016,
  author    = {Igor Poltavsky and Alexandre Tkatchenko},
  title     = {Modeling quantum nuclei with perturbed path integral molecular dynamics},
  journal   = {Chem. Sci.},
  year      = {2016},
  volume    = {7},
  number    = {2},
  pages     = {1368--1372},
  doi       = {10.1039/C5SC03443D},
  url       = {https://doi.org/10.1039/C5SC03443D},
  publisher = {Royal Society of Chemistry}
}

@article{Takahashi1984,
  author    = {Minoru Takahashi and Masatoshi Imada},
  title     = {{Monte Carlo} calculation of quantum systems. {II}. Higher order correction},
  journal   = {J. Phys. Soc. Jpn.},
  year      = {1984},
  volume    = {53},
  number    = {11},
  pages     = {3765--3769},
  doi       = {10.1143/JPSJ.53.3765},
  url       = {https://doi.org/10.1143/JPSJ.53.3765},
  publisher = {The Physical Society of Japan}
}

@article{Suzuki1995,
  author    = {Masuo Suzuki},
  title     = {Hybrid exponential product formulas for unbounded operators with possible applications to {Monte Carlo} simulations},
  journal   = {Phys. Lett. A},
  year      = {1995},
  volume    = {201},
  number    = {5--6},
  pages     = {425--428},
  doi       = {10.1016/0375-9601(95)00266-6},
  url       = {https://doi.org/10.1016/0375-9601(95)00266-6},
  publisher = {Elsevier}
}

@article{Chin1997,
  author    = {Siu A. Chin},
  title     = {Symplectic integrators from composite operator factorizations},
  journal   = {Phys. Lett. A},
  year      = {1997},
  volume    = {226},
  number    = {6},
  pages     = {344--348},
  doi       = {10.1016/S0375-9601(97)00003-0},
  url       = {https://doi.org/10.1016/S0375-9601(97)00003-0},
  publisher = {Elsevier}
}

@article{Kapil2016,
  author    = {Venkat Kapil and J{\"o}rg Behler and Michele Ceriotti},
  title     = {High-order path integrals made easy},
  journal   = {J. Chem. Phys.},
  year      = {2016},
  volume    = {145},
  number    = {23},
  pages     = {234103},
  doi       = {10.1063/1.4971438},
  url       = {https://doi.org/10.1063/1.4971438},
  publisher = {AIP Publishing}
}

@article{Herman1982,
  author    = {M. F. Herman and E. J. Bruskin and B. J. Berne},
  title     = {On path integral {Monte Carlo} simulations},
  journal   = {J. Chem. Phys.},
  year      = {1982},
  volume    = {76},
  number    = {10},
  pages     = {5150--5155},
  doi       = {10.1063/1.442815},
  url       = {https://doi.org/10.1063/1.442815},
  publisher = {AIP Publishing}
}

@article{Parrinello1984,
  author    = {Michele Parrinello and Aneesur Rahman},
  title     = {Study of an {F} center in molten {KCl}},
  journal   = {J. Chem. Phys.},
  year      = {1984},
  volume    = {80},
  number    = {2},
  pages     = {860--867},
  doi       = {10.1063/1.446740},
  url       = {https://doi.org/10.1063/1.446740},
  publisher = {AIP Publishing}
}

@article{Habershon2009,
  author    = {Scott Habershon and Thomas E. Markland and David E. Manolopoulos},
  title     = {Competing quantum effects in the dynamics of a flexible water model},
  journal   = {J. Chem. Phys.},
  year      = {2009},
  volume    = {131},
  number    = {2},
  pages     = {024501},
  doi       = {10.1063/1.3167790},
  url       = {https://doi.org/10.1063/1.3167790},
  publisher = {AIP Publishing}
}

@article{Hayward1997,
  author    = {Jennifer A. Hayward and Jeffrey R. Reimers},
  title     = {Unit cells for the simulation of hexagonal ice},
  journal   = {J. Chem. Phys.},
  year      = {1997},
  volume    = {106},
  number    = {4},
  pages     = {1518--1529},
  doi       = {10.1063/1.473300},
  url       = {https://doi.org/10.1063/1.473300},
  publisher = {AIP Publishing}
}

@article{Bussi2007,
  author    = {Giovanni Bussi and Davide Donadio and Michele Parrinello},
  title     = {Canonical sampling through velocity rescaling},
  journal   = {J. Chem. Phys.},
  year      = {2007},
  volume    = {126},
  number    = {1},
  pages     = {014101},
  doi       = {10.1063/1.2408420},
  url       = {https://doi.org/10.1063/1.2408420},
  publisher = {AIP Publishing}
}

@article{Trenins2019,
  author    = {George Trenins and Michael J. Willatt and Stuart C. Althorpe},
  title     = {Path-integral dynamics of water using curvilinear centroids},
  journal   = {J. Chem. Phys.},
  year      = {2019},
  volume    = {151},
  number    = {5},
  pages     = {054109},
  doi       = {10.1063/1.5100587},
  url       = {https://doi.org/10.1063/1.5100587},
  publisher = {AIP Publishing}
}

@article{Fletcher2021,
  author    = {Theo Fletcher and Andrew Zhu and Joseph E. Lawrence and David E. Manolopoulos},
  title     = {Fast quasi-centroid molecular dynamics},
  journal   = {J. Chem. Phys.},
  year      = {2021},
  volume    = {155},
  number    = {23},
  pages     = {231101},
  doi       = {10.1063/5.0076704},
  url       = {https://doi.org/10.1063/5.0076704},
  publisher = {AIP Publishing}
}

@article{Lawrence2023,
  author    = {Joseph E. Lawrence and Annina Z. Lieberherr and Theo Fletcher and David E. Manolopoulos},
  title     = {Fast quasi-centroid molecular dynamics for water and ice},
  journal   = {J. Phys. Chem. B},
  year      = {2023},
  volume    = {127},
  number    = {43},
  pages     = {9112--9123},
  doi       = {10.1021/acs.jpcb.3c05028},
  url       = {https://doi.org/10.1021/acs.jpcb.3c05028},
  publisher = {American Chemical Society}
}

@article{Musil2022,
  author    = {F{\'e}lix Musil and Iryna Zaporozhets and Frank No{\'e} and Cecilia Clementi and Venkat Kapil},
  title     = {Quantum dynamics using path integral coarse-graining},
  journal   = {J. Chem. Phys.},
  year      = {2022},
  volume    = {157},
  number    = {18},
  pages     = {181102},
  doi       = {10.1063/5.0120386},
  url       = {https://doi.org/10.1063/5.0120386},
  publisher = {AIP Publishing}
}

@misc{SCeffort,
  note = {To be fair to the projected Hessian {SC} method, Kapil {\em et al.}\ also
          introduced a multiple-time-step scheme in which the {SC} force on
          the even beads was evaluated only once every two time steps. This
          reduces the total number of force evaluations per time step from
          $3P/2$ to $5P/4$.}
}

@article{Cao1996,
  author    = {J. Cao and G. J. Martyna},
  title     = {Adiabatic path integral molecular dynamics methods. {II}. Algorithms},
  journal   = {J. Chem. Phys.},
  year      = {1996},
  volume    = {104},
  number    = {5},
  pages     = {2028--2035},
  doi       = {10.1063/1.470959},
  url       = {https://doi.org/10.1063/1.470959},
  publisher = {AIP Publishing}
}

@article{Ivanov2010,
  author    = {Sergei D. Ivanov and Ian M. Grant and Dominik Marx},
  title     = {Communications: On artificial frequency shifts in infrared spectra obtained from centroid molecular dynamics: Analytical insights and a simple cure},
  journal   = {J. Chem. Phys.},
  year      = {2010},
  volume    = {132},
  number    = {3},
  pages     = {031101},
  doi       = {10.1063/1.3290958},
  url       = {https://doi.org/10.1063/1.3290958},
  publisher = {AIP Publishing}
}

@article{Fan2022,
  author    = {Zheyong Fan and Yanzhou Wang and Penghua Ying and Keke Song and Junjie Wang and Yong Wang and Zezhu Zeng and Ke Xu and Eric Lindgren and J. Magnus Rahm and Alexander J. Gabourie and Jiahui Liu and Haikuan Dong and Jianyang Wu and Yue Chen and Zheng Zhong and Jian Sun and Paul Erhart and Yanjing Su and Tapio Ala-Nissila},
  title     = {{GPUMD}: A package for constructing accurate machine-learned potentials and performing highly efficient atomistic simulations},
  journal   = {J. Chem. Phys.},
  year      = {2022},
  volume    = {157},
  number    = {11},
  pages     = {114801},
  doi       = {10.1063/5.0106617},
  url       = {https://doi.org/10.1063/5.0106617},
  publisher = {AIP Publishing}
}

@article{Fan2021,
  author = {Zheyong Fan and Zezhu Zeng and Cunzhi Zhang and Yanzhou Wang and Keke Song and Haikuan Dong and Yue Chen and Tapio Ala-Nissila},
  title = {Neuroevolution machine learning potentials: Combining high accuracy and low cost in atomistic simulations and application to heat transport},
  journal = {Phys. Rev. B},
  volume = {104},
  number = {10},
  pages = {104309},
  year = {2021},
  publisher = {American Physical Society},
  doi = {10.1103/PhysRevB.104.104309},
  url = {https://doi.org/10.1103/PhysRevB.104.104309}
}

@misc{Zeng2026,
  author       = {Zezhu Zeng and David E. Manolopoulos},
  title        = {A patch to {GPUMD} for economised path-integrals},
  year         = {2026},
  publisher    = {GitHub},
  howpublished = {\url{https://github.com/ZengZezhu/Eco-Path-Integrals-in-GPUMD}},
  url          = {https://github.com/ZengZezhu/Eco-Path-Integrals-in-GPUMD},
}

@article{Lock2010,
  author    = {Nina Lock and Yue Wu and Mogens Christensen and Lisa J. Cameron and Vanessa K. Peterson and Adam J. Bridgeman and Cameron J. Kepert and Bo B. Iversen},
  title     = {Elucidating negative thermal expansion in {MOF-5}},
  journal   = {J. Phys. Chem. C},
  volume    = {114},
  number    = {39},
  pages     = {16181--16186},
  year      = {2010},
  publisher = {American Chemical Society},
  doi       = {10.1021/jp103212z},
  url       = {https://doi.org/10.1021/jp103212z}
}

@article{Kloutse2015,
  author    = {F. A. Kloutse and R. Zacharia and D. Cossement and R. Chahine},
  title     = {Specific heat capacities of {MOF-5}, {Cu-BTC}, {Fe-BTC}, {MOF-177} and {MIL-53 (Al)} over wide temperature ranges: Measurements and application of empirical group contribution method},
  journal   = {Microporous Mesoporous Mater.},
  volume    = {217},
  pages     = {1--5},
  year      = {2015},
  publisher = {Elsevier},
  doi       = {10.1016/j.micromeso.2015.05.047},
  url       = {https://doi.org/10.1016/j.micromeso.2015.05.047}
}

@article{Wieser2024,
  author  = {Sandro Wieser and Egbert Zojer},
  title   = {Machine learned force-fields for an {ab-initio} quality description of metal-organic frameworks},
  journal = {npj Comput. Mater.},
  volume  = {10},
  number  = {1},
  pages   = {18},
  year    = {2024},
  doi     = {10.1038/s41524-024-01205-w}
}

@article{Ying2025,
  author  = {Penghua Ying and Wenjiang Zhou and Lucas Svensson and Esm{\'e}e Berger and Erik Fransson and Fredrik Eriksson and Ke Xu and Ting Liang and Jianbin Xu and Bai Song and Shunda Chen and Paul Erhart and Zheyong Fan},
  title   = {Highly efficient path-integral molecular dynamics simulations with {GPUMD} using neuroevolution potentials: Case studies on thermal properties of materials},
  journal = {J. Chem. Phys.},
  volume  = {162},
  number  = {6},
  pages   = {064109},
  year    = {2025},
  doi     = {10.1063/5.0241006}
}

@article{Zeng2025,
  author  = {Zezhu Zeng and Zheyong Fan and Michele Simoncelli and Chen Chen and Ting Liang and Yue Chen and Geoff Thornton and Bingqing Cheng},
  title   = {Lattice distortion leads to glassy thermal transport in crystalline {Cs$_3$Bi$_2$I$_6$Cl$_3$}},
  journal = {Proc. Natl. Acad. Sci. U.S.A.},
  volume  = {122},
  number  = {41},
  pages   = {e2415664122},
  year    = {2025},
  doi     = {10.1073/pnas.2415664122}
}

@article{Goodfellow2026,
  author    = {Alister S. Goodfellow and Bao N. Nguyen},
  title     = {Graph-based internal coordinate analysis for transition state characterization},
  journal   = {J. Chem. Theory Comput.},
  year      = {2026},
  volume    = {22},
  number    = {5},
  pages     = {2348--2357},
  publisher = {American Chemical Society},
  doi       = {10.1021/acs.jctc.5c02073},
  url       = {https://doi.org/10.1021/acs.jctc.5c02073}
}

@article{Yamamoto2005,
  author    = {Toshiyuki Yamamoto},
  title     = {Path-integral virial estimator based on the scaling of fluctuation coordinates: Application to quantum clusters},
  journal   = {J. Chem. Phys.},
  year      = {2005},
  volume    = {123},
  number    = {10},
  pages     = {104101},
  doi       = {10.1063/1.2013257},
  url       = {https://doi.org/10.1063/1.2013257},
  publisher = {AIP Publishing}
}

@article{Lawrence2020,
  author    = {Joseph E. Lawrence and David E. Manolopoulos},
  title     = {Path integral methods for reaction rates in complex systems},
  journal   = {Faraday Discuss.},
  year      = {2020},
  volume    = {221},
  pages     = {9--29},
  doi       = {10.1039/C9FD00084D},
  url       = {https://doi.org/10.1039/C9FD00084D},
  publisher = {Royal Society of Chemistry}
}

@article{Craig2005,
  author    = {Ian R. Craig and David E. Manolopoulos},
  title     = {Chemical reaction rates from ring polymer molecular dynamics},
  journal   = {J. Chem. Phys.},
  year      = {2005},
  volume    = {122},
  number    = {8},
  pages     = {084106},
  doi       = {10.1063/1.1850093},
  url       = {https://doi.org/10.1063/1.1850093},
  publisher = {AIP Publishing}
}

@article{Craig2005b,
  author    = {Ian R. Craig and David E. Manolopoulos},
  title     = {A refined ring polymer molecular dynamics theory of chemical reaction rates},
  journal   = {J. Chem. Phys.},
  year      = {2005},
  volume    = {123},
  number    = {3},
  pages     = {034102},
  doi       = {10.1063/1.1954769},
  url       = {https://doi.org/10.1063/1.1954769},
  publisher = {AIP Publishing}
}

@article{Richardson2009,
  author    = {Jeremy O. Richardson and Stuart C. Althorpe},
  title     = {Ring-polymer molecular dynamics rate-theory in the deep-tunneling regime: Connection with semiclassical instanton theory},
  journal   = {J. Chem. Phys.},
  year      = {2009},
  volume    = {131},
  number    = {21},
  pages     = {214106},
  doi       = {10.1063/1.3267318},
  url       = {https://doi.org/10.1063/1.3267318},
  publisher = {AIP Publishing}
}

@article{Suleimanov2016,
  author    = {Yury V. Suleimanov and F. Javier Aoiz and Hua Guo},
  title     = {Chemical reaction rate coefficients from ring polymer molecular dynamics: Theory and practical applications},
  journal   = {J. Phys. Chem. A},
  year      = {2016},
  volume    = {120},
  number    = {43},
  pages     = {8488--8502},
  doi       = {10.1021/acs.jpca.6b07140},
  url       = {https://doi.org/10.1021/acs.jpca.6b07140},
  publisher = {American Chemical Society}
}

@article{Zhang2025,
  author    = {Liang Zhang and Florian Nitz and Dmitriy Borodin and Alec M. Wodtke and Hua Guo},
  title     = {Ring polymer molecular dynamics rates for hydrogen recombinative desorption on {Pt(111)}},
  journal   = {Precis. Chem.},
  year      = {2025},
  volume    = {3},
  number    = {6},
  pages     = {319--325},
  doi       = {10.1021/prechem.4c00099},
  url       = {https://doi.org/10.1021/prechem.4c00099},
  publisher = {American Chemical Society}
}

@article{Richardson2018,
  author    = {Jeremy O. Richardson},
  title     = {Perspective: Ring-polymer instanton theory},
  journal   = {J. Chem. Phys.},
  year      = {2018},
  volume    = {148},
  number    = {20},
  pages     = {200901},
  doi       = {10.1063/1.5028352},
  url       = {https://doi.org/10.1063/1.5028352},
  publisher = {AIP Publishing}
}

@article{Heller2020,
  author    = {Eric R. Heller and Jeremy O. Richardson},
  title     = {Instanton formulation of {Fermi}'s golden rule in the {Marcus} inverted regime},
  journal   = {J. Chem. Phys.},
  year      = {2020},
  volume    = {152},
  number    = {3},
  pages     = {034106},
  doi       = {10.1063/1.5137823},
  url       = {https://doi.org/10.1063/1.5137823},
  publisher = {AIP Publishing}
}

@article{Ansari2024,
  author    = {Imaad M. Ansari and Eric R. Heller and George Trenins and Jeremy O. Richardson},
  title     = {Heavy-atom tunnelling in singlet oxygen deactivation predicted by instanton theory with branch-point singularities},
  journal   = {Nat. Commun.},
  year      = {2024},
  volume    = {15},
  pages     = {4335},
  doi       = {10.1038/s41467-024-48463-2},
  url       = {https://doi.org/10.1038/s41467-024-48463-2},
  publisher = {Springer Nature}
}

@article{Castro2025,
  author    = {Jorge Castro and George Trenins and Venkat Kapil and Mariana Rossi},
  title     = {Vibrational spectra of materials and molecules from partially adiabatic elevated-temperature centroid molecular dynamics},
  journal   = {J. Chem. Phys.},
  year      = {2025},
  volume    = {163},
  number    = {20},
  pages     = {204102},
  doi       = {10.1063/5.0300048},
  url       = {https://doi.org/10.1063/5.0300048},
  publisher = {AIP Publishing}
}

\end{document}


\def\bra#1{\left<{#1}\right|}
\def\ket#1{\left|{#1}\right>}
\def\expval#1#2{\bra{#2} {#1} \ket{#2}}
\def\mapright#1{\smash{\mathop{\longrightarrow}\limits^{_{_{\phantom{X}}}{#1}_{_{\phantom{X}}}}}}

\title{Economised path integrals: Supplementary material}
\author{Zezhu Zeng}
\affiliation{Department of Chemistry, University of Oxford, Physical and Theoretical Chemistry Laboratory, South Parks Road, Oxford, OX1 3QZ, UK}
\author{David E. Manolopoulos}
\affiliation{Department of Chemistry, University of Oxford, Physical and Theoretical Chemistry Laboratory, South Parks Road, Oxford, OX1 3QZ, UK}

\begin{abstract}
{Section~A of this supplementary material contains a listing of the Fortran program we used to produce the results in Figs.~1--3 of the manuscript. Subroutine eco and its dependencies in this program provide a reference implementation of our Eco optimisation algorithm. We would strongly recommend the use of this reference implementation in preference to a canned Python (or other) non-linear least squares solver, which we have found to be less robust. In particular, we have not found a canned solver that gives a purely monotonic decrease in the rms fractional error in $R^2(\omega)$ over the full range of bead counts shown in our Fig.~1 (from $P=2$ to 100 in steps of 1).\\

\noindent
Section~B compares the results obtained using the ``pigs might fly" method described in Appendix A of the manuscript with fully adiabatic Te PIGS results\cite{Musil2022} for the dipole absorption spectrum of q-TIP4P/F ice at 150 K. The fully adiabatic Te PIGS results are also compared with partially adiabatic elevated temperature centroid molecular dynamics (PA-Te-CMD) results obtained using the method of Castro {\em et al.}\cite{Castro2025} Neither partially adiabatic method is perfect, but pigs might fly is found to perform slightly better than PA-Te-CMD by comparison with the Te PIGS benchmark. We suspect that the same will be true for the VDOS of ice at 100 K in Fig.~8 of the manuscript.}
\end{abstract}

\maketitle

\section*{A. Eco Fortran code}


\begin{verbatim}
      program main
      implicit real(8) (a-h,o-z)
c
c     ------------------------------------------------------------------ 
c     Minimises the rms fractional error in rG^2  
c     for a SHO with x in the range 0 < x < xmax.  
c     ------------------------------------------------------------------ 
c     
      allocatable :: y(:),yt(:),c(:),ct(:)
c
      read (5,*) xmax 
      pi = acos(-1.d0)
      nmax = nint(2*xmax)
      allocate (y(0:nmax-1),yt(0:nmax-1),c(0:nmax-1),ct(0:nmax-1))
c
      open (unit=10,file='error')
      open (unit=20,file='result')
      open (unit=30,file='coeffs')
      do n = 2,nmax
         y(0) = 0.d0
         call eco (xmax,n,y(1),rmst,rmsm,rmse)
         write (10,*) n,rmst,rmsm,rmse
         if (mod(n,(nmax/4)) .eq. 0) then
c
c           Trotter versus eco frequencies (in units of omega_P)
c
            write (20,"(1x,'# xmax = ',f6.2,' n = ',i4)") xmax,n
            yt(0) = 0.d0
            do k = 1,n-1
               yt(k) = 2*sin(k*pi/n)
               y(k) = y(k)/n
               write (20,*) k,yt(k),y(k)
            enddo
            write (20,*) '&'
c
c           Trotter versus eco circulant coefficients (dimensionless)
c
            ct = 0.d0
            ct(0) = 2.d0
            ct(1) = -1.d0
            ct(n-1) = -1.d0 
            call ytoc (y,c,n)
            sum = 0.d0
            do k = 0,n-1
               sum = sum-c(k)
            enddo
            write (30,"(1x,'# ',f9.4,i9,f18.10)") xmax,n,sum
            do k = 1,n-1
               write (30,*) k,ct(k),c(k)
            enddo
            write (30,*) '&'
         endif
      enddo
      close (unit=10)
      close (unit=20)
      close (unit=30)
c
      deallocate (y)
      end

      subroutine ytoc (y,c,n)
      implicit real(8) (a-h,o-z)
c
c     ------------------------------------------------------------------ 
c     Obtains the circulant coefficients c(0:n-1) from
c     a Fourier transform of the circulant eigenvalues 
c     y(0:n-1)^2.
c     ------------------------------------------------------------------ 
c
      dimension y(0:n-1),c(0:n-1)
c
      pibyn = acos(-1.d0)/n
      do l = 0,n-1
         c(l) = 0.d0
         do k = 0,n-1
            c(l) = c(l)+cos(2*pibyn*k*l)*y(k)**2
         enddo
         c(l) = c(l)/n
      enddo
      return
      end

      subroutine eco (xmax,n,y,rmst,rmsm,rmse)
      implicit real(8) (a-h,o-z)
      dimension y(n-1)
c
c     ------------------------------------------------------------------ 
c     Fits the dimensionless ring polymer frequencies 
c
c     y(k) = beta*hbar*omega(k) for k = 1,...,n-1 
c
c     to the radii of gyration of harmonic oscillators with
c     frequencies in the range 0 < beta*hbar*omega < xmax.
c
c     Returns the root mean square fractional errors in rG^2  
c     for the Trotter, Matsubara, and Eco approximations.
c     ------------------------------------------------------------------ 
c
      allocatable :: f(:),x(:),w(:),g(:),h(:,:)
      l = n/2
      m = nint(10*xmax)
      allocate (f(m),x(m),w(l),g(l),h(l,l))
c
c     Dense grid of oscillator frequencies between 0 and xmax 
c
      dx = xmax/m
      do j = 1,m
         x(j) = (j-0.5d0)*dx
         f(j) = x(j)**2/((x(j)/2)/tanh(x(j)/2)-1.d0)! 1/rG^2
      enddo
c
c     Fitting weights
c
      do k = 1,l
         w(k) = 2.d0
      enddo
      if (2*l .eq. n) w(l) = 1.d0 
c
c     Matsubara initial guess 
c
      pi = acos(-1.d0)
      do k = 1,l
         y(k) = 2*k*pi
      enddo
      call objfun (f,x,m,w,y,l,s,g,h)
      rmsm = sqrt(2*s)
c
c     Trotter initial guess
c
      do k = 1,l
         y(k) = 2*n*sin(k*pi/n)
      enddo
      call objfun (f,x,m,w,y,l,s,g,h)
      rmst = sqrt(2*s)
c
c     Minimisation
c
      iter = 0
      rmse = rmst
      print*
      print*,iter,rmse
      do iter = 1,10000
         sp = s
         call newton (f,x,m,w,y,l,s,g,h)
         rmse = sqrt(2*s)
         print*,iter,rmse
         if (s .ge. sp) exit
      enddo
c
c     Return y(k) for k = 1,n-1
c
      do k = 1,(n-1)/2
         y(n-k) = y(k)
      enddo
      deallocate (f,x,w,g,h)
      return
      end   

      subroutine newton (f,x,m,w,y,l,s,g,h)
      implicit real(8) (a-h,o-z)
      dimension f(m),x(m),w(l),y(l),g(l),h(l,l)
c
c     ------------------------------------------------------------------ 
c     Single safe Newton step to minimise s(y).
c     y, s, g, and h are all updated on return.
c     ------------------------------------------------------------------  
c
      allocatable :: z(:),dy(:)
      allocate (z(l),dy(l))
c
c     Calculate the Newton shift dy and check it is a descent direction
c
      call shift (g,h,l,dy)
      s0 = s
      g0 = 0.d0
      do j = 1,l
         g0 = g0+g(j)*dy(j)
      enddo
      if (g0 .gt. 0.d0) stop 'dy is not a descent direction in newton'
c
c     Apply a full or partial Newton shift that both decreases s(y)
c     and keeps the y(j)'s in ascending order
c
      c = 2.d0
      do iter = 1,60
         c = c/2
         z = y+c*dy
         if (z(1) .lt. 0.d0) go to 1
         do j = 2,l
            if (z(j) .lt. z(j-1)) go to 1
         enddo
         call objfun (f,x,m,w,z,l,s,g,h)
         if (s .le. s0) go to 2
   1  enddo 
      stop 'failed to go downhill in newton'
   2  y = z
      deallocate (z,dy)
      return
      end

      subroutine shift (g,h,l,dy)
      implicit real(8) (a-h,o-z)
      dimension g(l),h(l,l),dy(l)
c
c     ------------------------------------------------------------------ 
c     Calculates the Newton shift dy = -H^{-1}*g.
c     ------------------------------------------------------------------ 
c
      allocatable :: d(:),f(:)
      allocate (d(l),f(l))
c
      call symevp (h,l,l,d,ierr)
      if (ierr .ne. 0) stop 'symevp failed in shift'
      dy = matmul(transpose(h),g)
      delta = max(1.d-16*d(l),-2*d(1))
      do j = 1,l
         f(j) = -dy(j)/(d(j)+delta)
      enddo
      dy = matmul(h,f)
      deallocate (d,f)
      return
      end

      subroutine objfun (f,x,m,w,y,l,s,g,h)
      implicit real(8) (a-h,o-z)
      dimension f(m),x(m),w(l),y(l),g(l),h(l,l)
c
c     ------------------------------------------------------------------ 
c     Calculates 
c
c     s(y) = (1/2m) sum_{j=1}^m r(j)^2
c
c     where
c
c     r(j) = f(j)*sum_{k=1}^l w(k)/[x(j)^2+y(k)^2] - 1,
c
c     and its gradient g(y) and hessian H(y).
c     ------------------------------------------------------------------ 
c
      allocatable :: dg(:),dh(:)
      allocate (dg(l),dh(l))
c
      s = 0.d0 
      g = 0.d0
      h = 0.d0 
      do j = 1,m
         x2 = x(j)**2
         r = -1.d0
         do k = 1,l
            d = 1.d0/(y(k)**2+x2)
            e = f(j)*w(k)*d
            r = r+e
            dg(k) = -2*d*e*y(k)
            dh(k) = 2*d*d*e*(3*y(k)**2-x2)
         enddo
         s = s+r**2/2
         do k = 1,l
            g(k) = g(k)+r*dg(k)
            do i = 1,k
               h(i,k) = h(i,k)+dg(i)*dg(k)
            enddo
            h(k,k) = h(k,k)+r*dh(k)
         enddo
      enddo
      s = s/m
      g = g/m
      h = h/m
      deallocate (dg,dh)
      return
      end

      subroutine symevp (a,lda,n,d,ierr)
      implicit real(8) (a-h,o-z)
      dimension a(lda,n),d(n)
c
c     ----------------------------------------------------------------- 
c     Uses LAPACK DSYEV to diagonalise a real symmetric matrix.
c     ----------------------------------------------------------------- 
c
      allocatable :: work(:)
c
      lwork = 34*n
      allocate (work(lwork))
      call dsyev ('V','U',n,a,lda,d,work,lwork,ierr)
      deallocate (work)
      return
      end
\end{verbatim}

\newpage
\section*{B. Dipole absorption spectra}

\begin{figure}[h]
    \centering
    \resizebox{0.45\columnwidth}{!}{\includegraphics{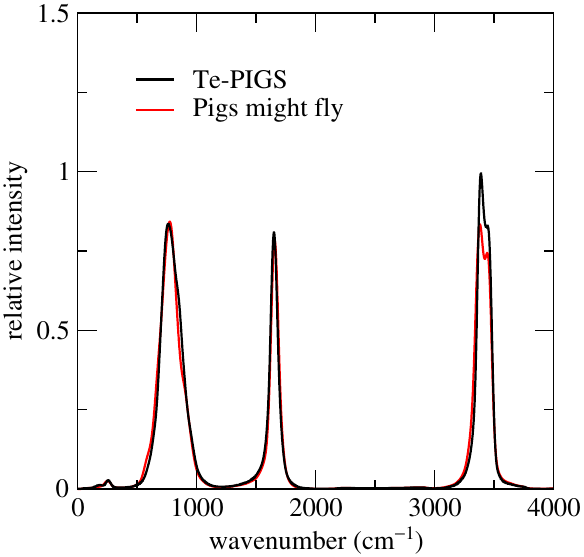}}
    \resizebox{0.45\columnwidth}{!}{\includegraphics{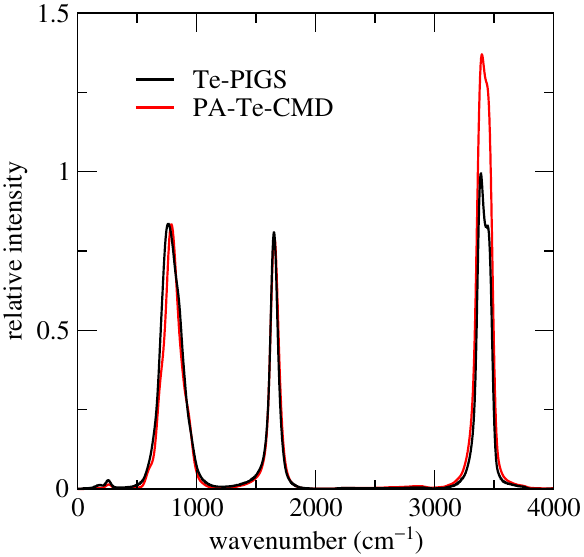}}
    \caption{Comparison of fully adiabatic Te PIGS, ``pigs might fly", and PA-Te-CMD dipole absorption spectra of q-TIP4P/F ice at 150 K. The Te PIGS spectrum was taken from Musil {\em et al.}~\cite{Musil2022} The pigs might fly spectrum was calculated using the same parameters as the VDOS spectrum in Fig.~8 of the present manuscript ($T_e=600$ K and $\Omega=40,000$ cm$^{-1}$). The PA-Te-CMD spectrum was calculated using the same parameters as in Fig.~S12 of Castro {\em et al.}'s supplementary material\cite{Castro2025} ($T_e=500$ K, $\Omega=24,000$ cm$^{-1}$, $\tau=10$ fs, and $\lambda=0.001$). The main difference between the two partially adiabatic (red) calculations is that PA-Te-CMD overestimates the intensity of the O--H stretching band whereas ``pigs might fly" underestimates it.}
    \label{fig:S1}
\end{figure}

\bibliography{ecorefs.bib}